\documentclass[seceq]{ptptex}

\usepackage{graphicx}
\usepackage{wrapft}
\usepackage{amsmath,amssymb}
\usepackage{bm}
\usepackage{color}
\usepackage{textcomp}

\notypesetlogo

\title{Quadrupole Deformation $\beta$ and $\gamma$ Constraint 
in a Framework of Antisymmetrized Molecular Dynamics}

\author{Tadahiro \textsc{Suhara}$^{1}$ and
Yoshiko \textsc{Kanada-En'yo}$^{2}$}

\inst{$^{1}$Department of Physics, Graduate School of Science, Kyoto University, Kyoto 606-8502, Japan \\
$^{2}$Yukawa Institute for Theoretical Physics, Kyoto University, Kyoto 606-8502, Japan}

\abst{We propose a new method of the $\beta$-$\gamma$ constraint for quadrupole deformation
in antisymmetrized molecular dynamics (AMD)
to describe various cluster and shell-model structures in the ground and excited states of 
light nuclei.
We apply this method to $N=6$ isotones, $^{10}$Be, $^{12}$C, $^{9}$Li, and $^{11}$B, and find 
various structures as functions of the deformation parameters, $\beta$ and $\gamma$.
In these nuclei, shell-model-like structures appear in the small $\beta$ region, 
while cluster structures develop well in the large $\beta$ region where 
various geometric configurations of clusters are obtained depending on the $\gamma$ parameter.
For $^{10}$Be and $^{12}$C, we superpose the basis AMD wave functions obtained by the $\beta$-$\gamma$ constraint method 
to calculate energy spectra, and prove the advantages of the present method of the two-dimensional $\beta$-$\gamma$ constraint 
in the framework of AMD.}

\begin{document}

\maketitle

\section{Introduction}\label{introduction}

In light nuclei, the cluster aspect is one of the essential features, as well as the shell-model aspect.
Owing to the coexistence of these two natures, namely, cluster and shell-model features, various structures 
appear in stable and unstable nuclei.

$^{12}$C is one of the typical examples where the cluster and shell-model aspects coexist.
The ground state of $^{12}$C is known to have mainly a shell-model feature of 
the $p_{3/2}$ subshell closed configuration, whereas
the well-developed 3$\alpha$-cluster structures appear in excited states.
In the theoretical works on the 3$\alpha$-cluster structures,\cite{Horiuchi_OCM_74,Uegaki_12C_77,Kamimura_12C_77,Descouvemont_12C_87,En'yo_12C_98,Tohsaki_12C_01,Funaki_12C_03,Neff_12C_04,En'yo_12C_07,Kurokawa_12C_07} \ 
various configurations of the 3$\alpha$-cluster structures were suggested in the excited states
above the 3$\alpha$ threshold energy, for example, 
the $\alpha$ condensation of weakly interacting 
three $\alpha$ clusters in the $0^{+}_2$ state and 
the equilateral-triangular structure of three $\alpha$ clusters in the $3^{-}_{1}$ state.
Moreover, a linear-chainlike (or an obtuse-angle-triangular) structure 
of three $\alpha$ clusters in the $0^{+}_{3}$ state was suggested.

Cluster structures have also been found in light neutron-rich nuclei such as Be isotopes.
In $^{10}$Be, the low-lying states are understood in a molecular
2$\alpha+2n$ picture,\cite{vonOertzen_ClusterRev_06,Itagaki_10Be_00} \ where 
two $\alpha$ cores are formed and two excess neutrons occupy molecular orbitals around the $2\alpha$.
In terms of a simple shell model, $^{10}$Be is an $N=6$ nucleus, and therefore, the $p_{3/2}$ subshell closure 
effect is also important, as well as the $2\alpha+2n$ cluster feature at least in the ground state. 
This means that the cluster-shell competition is essential in unstable nuclei 
as well as stable nuclei, as argued in Ref.~\citen{Itagaki_ClusterShellCompetition_04}.

For theoretical investigations of such nuclei, it is necessary
to describe the coexistence of shell and cluster features systematically. 
However, many theoretical frameworks still have deficiencies 
in describing both the shell-model and cluster structures.
In fact, in the case of $^{12}$C, shell models can be used to describe low-lying shell-model states but they
usually fail to describe high-lying 3$\alpha$-cluster states. 
On the other hand, conventional cluster models are suitable for studying the 3$\alpha$-cluster states,
but it is not easy to reproduce well the detailed properties of 
low-lying shell-model states because $\alpha$ cluster breaking is not incorporated
in the cluster models.

A method of antisymmetrized molecular dynamics (AMD)\cite{En'yo_PTP_95,En'yo_AMD_95} \ is one of the frameworks useful for overcoming this
problem. It was applied to $^{12}$C and 
succeeded to describe the shell and cluster features due to the flexibility of its 
wave functions.\cite{En'yo_12C_98,En'yo_12C_07} \ 
Moreover, in the study of fermionic molecular dynamics (FMD), in which model wave functions 
are similar to those of AMD, the coexistence of shell and cluster features in $^{12}$C was 
described successfully.\cite{Neff_12C_04} \ 

The AMD method has also been applied to various stable and unstable nuclei, 
and it has been proved to be one of the powerful approaches of describing various structures
such as cluster structures and shell-model structures.\cite{En'yo_AMD_95,En'yo_AMD_03,En'yo_sup_01} \ 
There are some versions of the AMD, for example, the variation after parity and total-angular-momentum projections (VAPs),\cite{En'yo_12C_98} \ 
the variation with the constraint on the quadrupole deformation 
$\beta$ ($\beta$ constraint AMD),\cite{Dote_Beta-Constraint_97,Kimura_uptoMg_01,En'yo_AMD_03} \ or the constraint on 
the cluster distances ($d$-constraint AMD).\cite{Taniguchi_D-Constraint_04} \ 
In principle, a basis AMD wave function is given by a Slater determinant of Gaussian wave packets, and
excited states are described by superposition of Slater determinants.
In practical calculations of excited states of light nuclei, it is important to prepare efficiently 
various cluster configurations 
including 2-body and 3-body clusterings as basis wave functions in the AMD framework.
Moreover, in the study of unstable nuclei, further flexible model wave functions such as 
2-body or 3-body cluster structures with surrounding valence nucleons will be required to describe 
possible exotic cluster structures in excited states.

To study a variety of cluster structures and the coexistence of cluster and shell features
in light unstable nuclei,
we propose an extended method of constraint AMD
to describe various cluster and shell structures. That is the
two-dimensional constraint with respect to the quadrupole deformation parameters, $\beta$ and $\gamma$,
which is expected to be efficient for preparing basis wave functions with various cluster configurations.
We call this method $\beta$-$\gamma$ constraint AMD.
We expect shell-model structures to appear in the small $\beta$ region, whereas
developed 2-body or 3-body cluster structures can be obtained for large $\beta$.
In the large $\beta$ region, various configurations of cluster structures may appear
depending on $\beta$ and $\gamma$. 

The $\beta$-$\gamma$ constraint AMD may also be useful
in the study of triaxial deformations.
On the other hand, in the Hartree-Fock-Bogolyubov (HFB) calculations, 
the $\beta$-$\gamma$ constraint was adopted, for example, 
in Ref.~\citen{Girod_triaxial_83}, and 
the superposition of $\beta$-$\gamma$ constraint wave functions
has been performed recently by Bender and Heenen.\cite{Bender_24Mg_08}.
It was found that the triaxiality is important to reproduce the experimental 
data of $^{24}$Mg in the HFB calculations. 
However, works on triaxial calculations with the superposition are limited, and
it is still a challenging problem.
Moreover, such mean-field approaches are not necessarily 
suitable for describing cluster structures.
For the study of cluster features, it is important to apply 
the $\beta$-$\gamma$ constraint to a framework that can describe cluster structures.

In this paper, we applied the $\beta$-$\gamma$ constraint AMD
to $N=6$ isotones, $^{10}$Be, $^{12}$C, $^{9}$Li, and $^{11}$B
to check the applicability of this method.
We analyze the results and confirm that various structures appear as functions of 
the deformation parameters, $\beta$ and $\gamma$, in the present framework. 
In particular, we focus on the coexistence of shell and cluster features. 
For $^{10}$Be and $^{12}$C, we also calculate the energy spectra of excited states
by the superposition of the obtained basis wave functions and compare the results with 
the experimental data.
We show that the $\beta$-$\gamma$ constraint AMD is useful for reproducing the energy spectra.
A role of the $\gamma$ degree of freedom is also discussed.

The content of this paper is as follows. 
In \S \ref{framework}, we explain the framework of the $\beta$-$\gamma$ constraint AMD.
The calculated results are shown in \S \ref{results}.
In \S \ref{discussions}, we discuss the effect of the triaxial deformation parameter $\gamma$.
Finally, in \S \ref{summary}, a summary and an outlook are given.

\section{Framework of $\beta$-$\gamma$ constraint AMD}\label{framework}

We adopt a method of AMD with constraint. 
The frameworks of AMD and constraint AMD are described in detail, for example, in 
Refs.~\citen{En'yo_AMD_95,En'yo_sup_01,En'yo_AMD_03}.
In this paper, we propose a two-dimensional constraint with respect to quadrupole deformation
parameters.

\subsection{Wave function of AMD}

In the method of AMD, 
a basis wave function of an $A$-nucleon system $|\Phi \rangle$ 
is described by a Slater determinant of single-particle wave functions $|\varphi_{i} \rangle$ as
\begin{equation}
|\Phi \rangle = \frac{1}{\sqrt{A!}} \det \left\{ |\varphi_{1} \rangle, \cdots ,|\varphi_{A} \rangle \right\}.
\end{equation}
The $i$-th single-particle wave function $|\varphi_{i} \rangle$ consists of 
the spatial part $|\phi_{i} \rangle$, spin part $|\chi_{i} \rangle$, and isospin part $|\tau_{i} \rangle$ as
\begin{equation}
	|\varphi_{i} \rangle = |\phi_{i} \rangle |\chi_{i} \rangle |\tau_{i} \rangle.
\end{equation}
The spatial part $|\phi_{i} \rangle$ is given by a Gaussian wave packet
whose center is located at $\bm{Z}_{i}/\sqrt{\nu}$ as
\begin{equation}
	\langle \bm{r} | \phi_{i} \rangle = \left( \frac{2\nu}{\pi} \right)^{\frac{3}{4}}
		\exp \left[ - \nu \left( \bm{r} - \frac{\bm{Z}_{i}}{\sqrt{\nu}} \right)^{2} 
		+ \frac{1}{2} \bm{Z}_{i}^{2}\right] 
	\label{single_particle_spatial}, 
\end{equation}
where $\nu$ is the width parameter and is taken to be a common value for all the
single-particle Gaussian wave functions in the present work.
The spin orientation is given by the parameter $\bm{\xi}_{i}$, while
the isospin part $|\tau_{i} \rangle$ is fixed to be up (proton) or down (neutron), 
\begin{align}
	|\chi_{i} \rangle &= \xi_{i\uparrow} |\uparrow \ \rangle + \xi_{i\downarrow} |\downarrow \ \rangle,\\
	|\tau_{i} \rangle &= |p \rangle \ or \ |n \rangle.
\end{align}
In a basis wave function $|\Phi \rangle$, $\{ X \} \equiv \{ \bm{Z} , \bm{\xi} \} = \{ \bm{Z}_{1} , \bm{\xi}_{1} , \bm{Z}_{2} , \bm{\xi}_{2} , 
\cdots , \bm{Z}_{A} , \bm{\xi}_{A} \}$ are complex variational parameters and they 
are determined by the energy optimization using the frictional cooling method.\cite{En'yo_sup_01,En'yo_AMD_03} \ 
As the variational wave function, we employ the parity-projected wave function
\begin{equation}
	|\Phi ^{\pm} \rangle = P^{\pm} |\Phi \rangle = \frac{1 \pm P}{2} |\Phi \rangle.
\end{equation}
Here, $P$ is the parity transformation operator. 
We perform the variation for the parity-projected energy
$\langle \Phi ^{\pm}| H |\Phi ^{\pm} \rangle / \langle \Phi ^{\pm}|\Phi ^{\pm} \rangle$,
where $H$ is the Hamiltonian.
After the variation, we project the obtained wave function onto the 
total-angular-momentum eigenstate.
It means that the parity projection is performed before the variation, and
the total-angular-momentum projection is carried after the variation.

\subsection{$\beta$-$\gamma$ constraint}

To describe various cluster and shell-model structures that may appear 
in the ground and excited states of light nuclei,
we constrain the quadrupole deformation parameters, $\beta$ and $\gamma$, and perform the 
energy variation with the constraints on the $\beta$-$\gamma$ plane.

The deformation parameters, $\beta$ and $\gamma$, are defined as
\begin{align}
	&\beta \cos \gamma \equiv \frac{\sqrt{5\pi}}{3} 
		\frac{2\langle z^{2} \rangle -\langle x^{2} \rangle -\langle y^{2} \rangle }{R^{2}}, \\
	&\beta \sin \gamma \equiv \sqrt{\frac{5\pi}{3}} 
		\frac{\langle x^{2} \rangle -\langle y^{2} \rangle }{R^{2}} \label{definition_beta_gamma}, \\
	&R^{2} \equiv \frac{5}{3} \left( \langle x^{2} \rangle + \langle y^{2} \rangle 
		+ \langle z^{2} \rangle \right).
\end{align}
Here, $\langle O \rangle$ represents the expectation value of the operator $O$ for an intrinsic wave function $| \Phi \rangle$.
$x$, $y$, and $z$ are the inertia principal axes that are chosen as
$\langle y^{2} \rangle \le \langle x^{2} \rangle \le \langle z^{2} \rangle $ and
$\langle xy \rangle = \langle yz \rangle = \langle zx \rangle =0$.
To satisfy the latter condition, we also impose 
the constraints $\langle xy \rangle/R^{2} = \langle yz \rangle/R^{2} = \langle zx \rangle/R^{2} =0$. 
To obtain the energy minimum state under the constraint condition,
we add the constraint potential $V_{\text{const}}$ to the total energy of the system
in the energy variation. The constraint potential $V_{\text{const}}$ is given as
\begin{align} 
	V_{\text{const}} \equiv &\eta_{1} 
	\left[ (\beta \cos \gamma - \beta_{0} \cos \gamma_{0})^{2} + (\beta \sin \gamma - \beta_{0} \sin \gamma_{0})^{2} \right] \notag \\
	+ &\eta_{2} \left[ \left( \frac{\langle xy \rangle}{R^{2}} \right)^{2} 
		+ \left( \frac{\langle yz \rangle}{R^{2}} \right)^{2} 
		+ \left( \frac{\langle zx \rangle}{R^{2}} \right)^{2} \right].
	\label{constraint_energy}
\end{align}
Here, $\eta_{1}$ and $\eta_{2}$ take sufficiently large values.
After the variation with the constraint, we obtain the optimized wave functions
$|\Phi^{\pm}(\beta_{0}, \gamma_{0}) \rangle$
for each set of parameters, $(\beta, \gamma) = (\beta_{0}, \gamma_{0})$.

In the calculations of energy levels, 
we superpose the total-angular-momentum projected 
wave functions $P^{J}_{MK} |\Phi^{\pm}(\beta, \gamma) \rangle$. 
Thus, the final wave function for the $J^\pm_n$ state is given by
a linear combination of the basis wave functions as 
\begin{equation}
	|\Phi ^{J\pm}_{n} \rangle = \sum_{K} \sum_{i} f_{n}(\beta_{i}, \gamma_{i}, K) P^{J}_{MK} |\Phi^{\pm}(\beta_{i}, \gamma_{i}) \rangle.
	\label{dispersed_GCM}
\end{equation}
The coefficients $f_{n}(\beta_{i}, \gamma_{i}, K)$ are determined using the Hill-Wheeler equation
\begin{equation}
	\delta \left( \langle \Phi ^{J\pm}_{n} | H | \Phi ^{J\pm}_{n} \rangle - 
	E_{n} \langle \Phi ^{J\pm}_{n} | \Phi ^{J\pm}_{n} \rangle\right) = 0.
	\label{Hill-Wheeler}
\end{equation}
This means the superposition of multiconfigurations described by 
parity and total-angular-momentum projected AMD wave functions.
In the limit of sufficient basis wave functions 
on the $\beta$-$\gamma$ plane, it corresponds to the
generator coordinate method (GCM) with the two-dimensional generator coordinates 
of the quadrupole deformation parameters, $\beta$ and $\gamma$.

\subsection{Hamiltonian and parameters}

The Hamiltonian $H$ consists of the kinetic term 
and effective two-body interactions as
\begin{equation}
	H = \sum_{i} t_{i} - T_{\text{G}} + \sum_{i<j} V^{\text{central}}_{ij} 
	+ \sum_{i<j} V^{\text{spin-orbit}}_{ij} + \sum_{i<j} V^{\text{Coulomb}}_{ij},
\end{equation}
where $V^{\text{central}}_{ij}$, $V^{\text{spin-orbit}}_{ij}$ and $V^{\text{Coulomb}}_{ij}$ 
are the central force, spin-orbit force, and Coulomb force.
As the central force, we use the Volkov No.~2 interaction,\cite{Volkov_No2_65} \ 
\begin{equation}
	V^{\text{central}}_{ij} = \sum_{k=1}^{2} v_{k} \exp \left[- \left( \frac{r_{ij}}{a_{k}} \right)^{2} \right]
                                  (W + B P_{\sigma} - H P_{\tau} - M P_{\sigma} P_{\tau}),
\end{equation}
where $v_{1} = -60.65$ MeV, $v_{2} = 61.14$ MeV, $a_{1} = 1.80$ fm, and $a_{2} = 1.01$ fm.
For the spin-orbit part, we used the spin-orbit term of the G3RS interaction,\cite{G3RS_79} \ which is a two-range
Gaussian with a projection operator $P(^{3} \mbox{O})$ onto the triplet odd state,
\begin{align}
	V^{\text{spin-orbit}}_{ij} = \sum_{k=1}^{2} &u_{k} \exp \left[- \left( \frac{r_{ij}}{b_{k}} \right)^{2} \right] P(^{3} \mbox{O}) \bm{L} \cdot \bm{S}, \\
	P(^{3} \mbox{O}) &= \frac{1+P_{\sigma}}{2} \frac{1+P_{\tau}}{2},
\end{align}
where $b_{1} = 0.600$ fm and $b_{2} = 0.477$ fm.

We take interaction parameters, the Majorana exchange parameter $M = 0.6$ ($W = 0.4$), the Bartlett exchange parameter $B = 0.125$,
and the Heisenberg exchange parameter $H = 0.125$ in the central force, and 
$u_{1} = -1600$ MeV and $u_{2} = 1600$ MeV in the spin-orbit force.
All these parameters are 
the same as those adopted in Refs.~\citen{Itagaki_10Be_00} and \citen{Okabe_parameter_79}, 
except for a small modification in the strength of the spin-orbit force.
They are adjusted to reproduce 
the $\alpha + \alpha$ phase shift ($M$, $W=1-M$, $a_{1}$, $a_{2}$), binding energy of the deuteron ($B=H$),
and $\alpha + n$ phase shift ($u_{1}=-u_{2}$, $b_{1}$, $b_{2}$).
We slightly modify only the strengths of the spin-orbit force 
from $-u_{1} = u_{2} = 2000$ MeV adopted in Ref.~\citen{Itagaki_10Be_00}
to  $-u_{1} = u_{2} = 1600$ MeV to fit the $0^{+}_{1}$ energy of $^{12}$C 
in the present calculations.
By using these parameters, 
the deuteron binding energy in an exact calculation is 2.24 MeV, which
is in agreement with the experimental value, 2.22 MeV.
When the deuteron wave function is approximated by one AMD wave function, 
the calculated binding energy is 0.67 MeV, which slightly underestimates the experimental 
value, 2.2 MeV.
The Coulomb force $V^{\text{Coulomb}}_{ij}$ is approximated by a sum of seven Gaussians.

For the width parameter of
single-particle Gaussian wave packets in Eq.~\eqref{single_particle_spatial}, 
we used the value $\nu = 0.235$ fm$^{-2}$, which is 
determined from a variational calculation 
for the ground state of $^{9}$Be in Ref.~\citen{Okabe_parameter_79}.

\section{Results}\label{results}

We applied the $\beta$-$\gamma$ constraint 
AMD to the $N=6$ isotones, $^{10}$Be, $^{12}$C, $^9$Li, and $^{11}$B.
In this section, we analyze the results of the $\beta$-$\gamma$ constraint
and show that 
various structures including well-developed cluster ones appear as functions of 
deformation parameters, $\beta$ and $\gamma$, in the present framework. 
In particular, we focus on the coexistence of shell and cluster features. 
For $^{10}$Be and $^{12}$C, we also show the energy spectra of the excited states
calculated by the superposition of the basis wave functions and compare them with the experimental data.

\subsection{Energy surfaces}

\begin{figure}[tb]
	\parbox{\halftext}{\includegraphics[width=6.6cm, bb=10 12 693 323, clip]{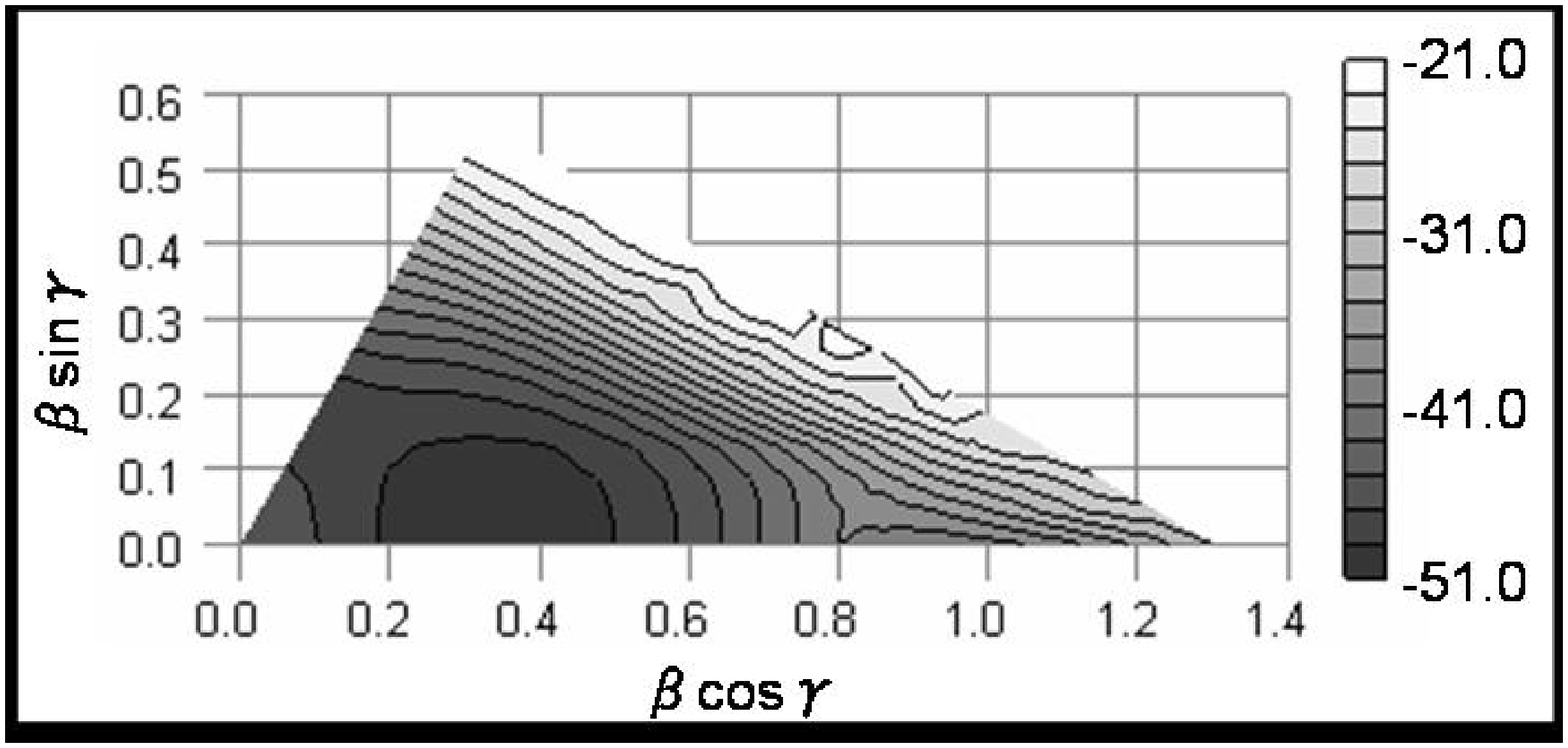}}
	\hfill
	\parbox{\halftext}{\includegraphics[width=6.6cm, bb=10 12 693 323, clip]{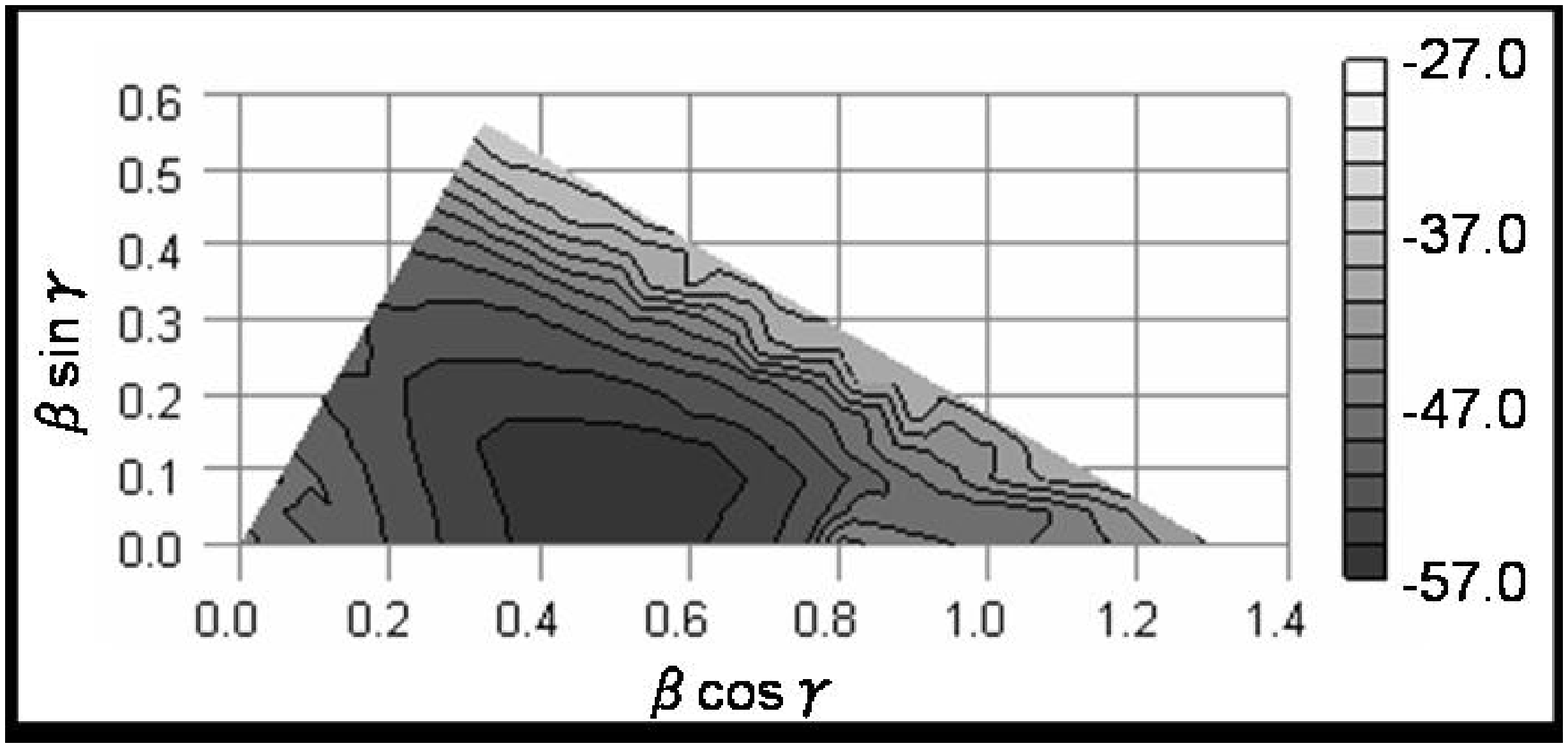}}
	\caption{Energy surfaces of $^{10}$Be on the $\beta$-$\gamma$ plane.
	The left shows the energy for the positive-parity states before the total-angular-momentum projection and 
	the right shows that for the $0^{+}$ states after the total-angular-momentum projection.}
	\label{10Be_energy_surface}
\end{figure}

\begin{figure}[tb]
	\parbox{\halftext}{\includegraphics[width=6.6cm, bb=10 12 693 323, clip]{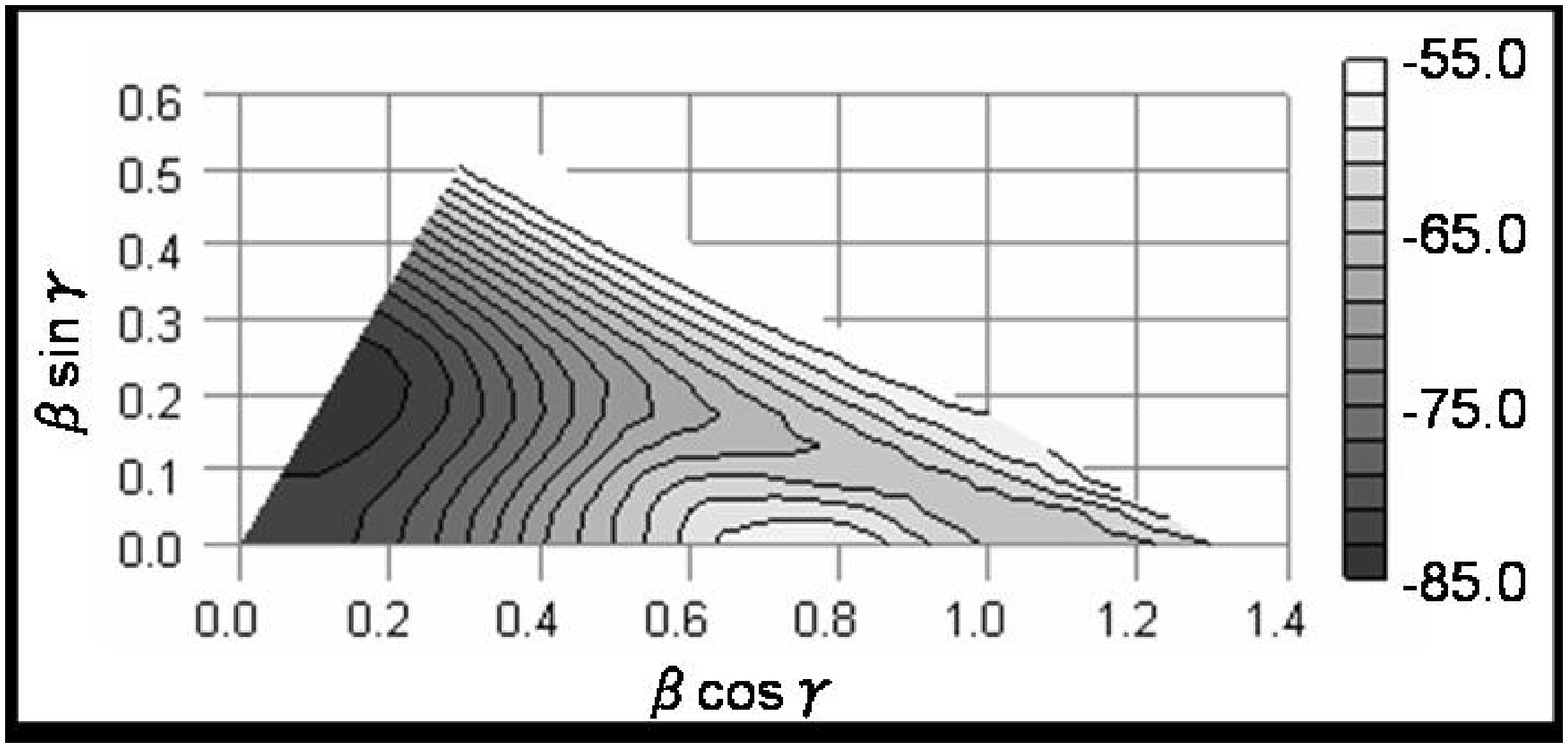}}
	\hfill
	\parbox{\halftext}{\includegraphics[width=6.6cm, bb=10 12 693 323, clip]{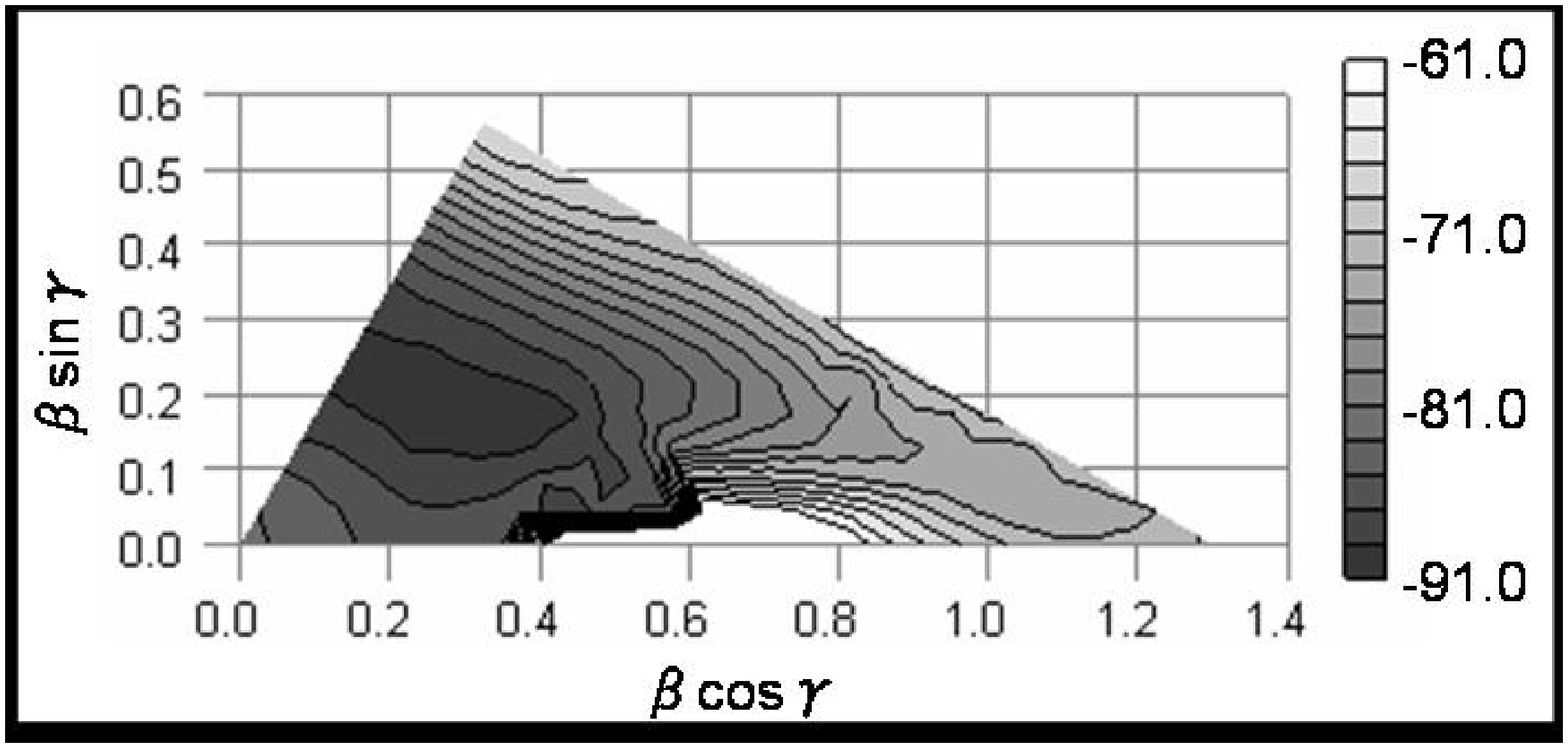}}
	\caption{Energy surfaces of $^{12}$C on the $\beta$-$\gamma$ plane.
	The left shows the energy for the positive-parity states and 
	the right shows that for the $0^{+}$ states after the total-angular-momentum projection.}
	\label{12C_energy_surface}
\end{figure}

\begin{figure}[tb]
	\parbox{\halftext}{\includegraphics[width=6.6cm, bb=10 12 693 323, clip]{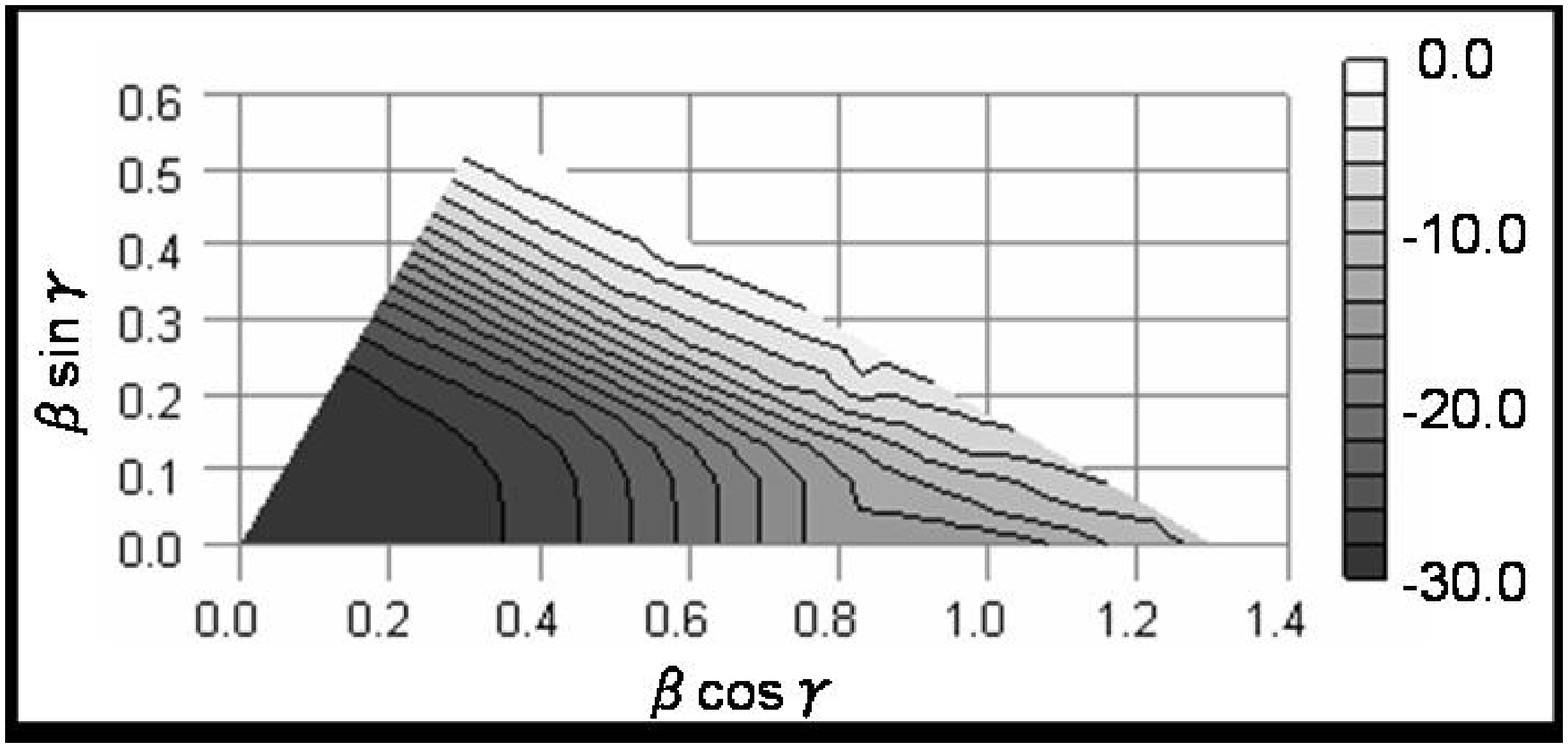}}
	\hfill
	\parbox{\halftext}{\includegraphics[width=6.6cm, bb=10 12 693 323, clip]{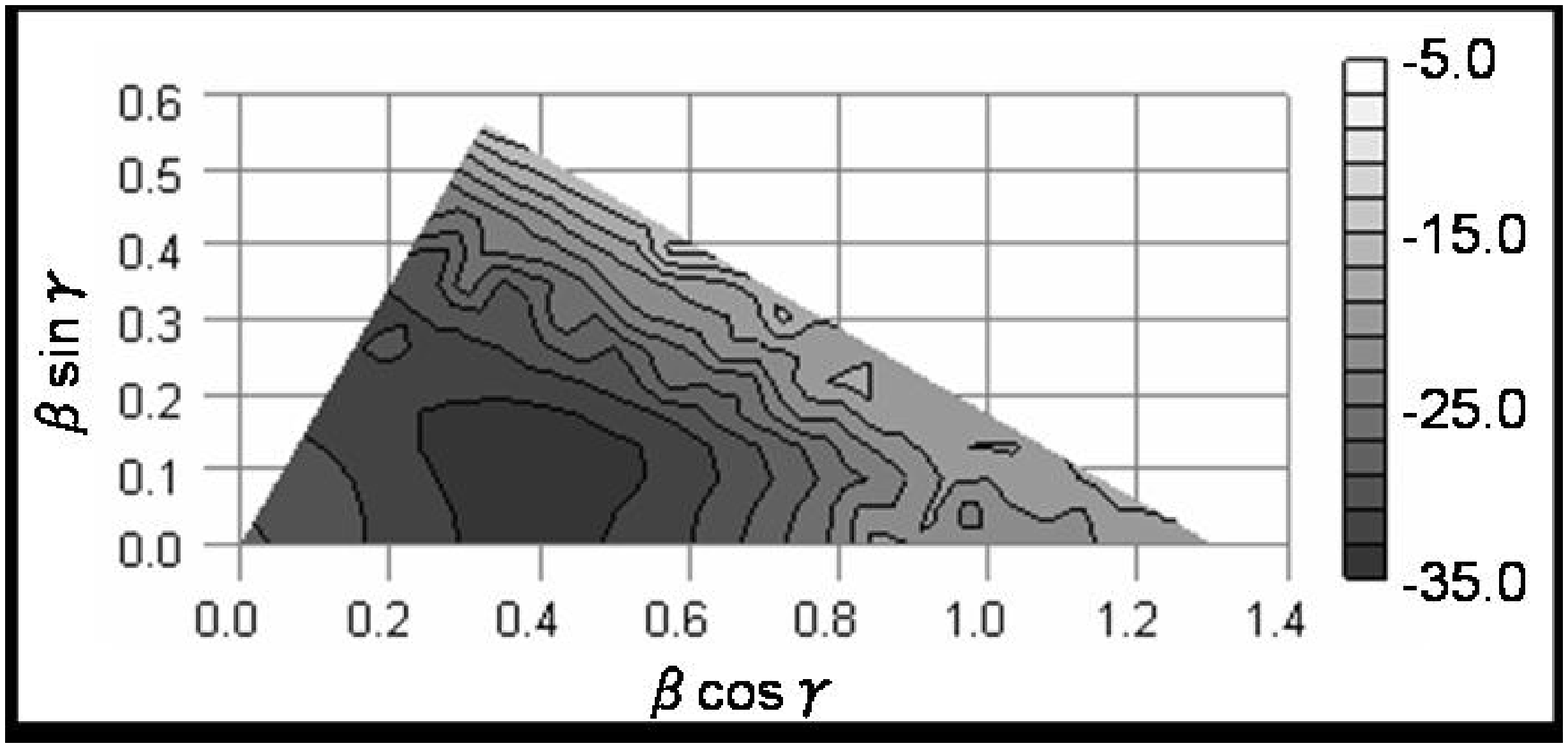}}
	\caption{Energy surfaces of $^{9}$Li on the $\beta$-$\gamma$ plane.
	The left shows the energy for the negative-parity states and 
	the right shows that for the $3/2^{-}$ states after the total-angular-momentum projection.}
	\label{9Li_energy_surface}
\end{figure}

\begin{figure}[tb]
	\parbox{\halftext}{\includegraphics[width=6.6cm, bb=10 12 693 323, clip]{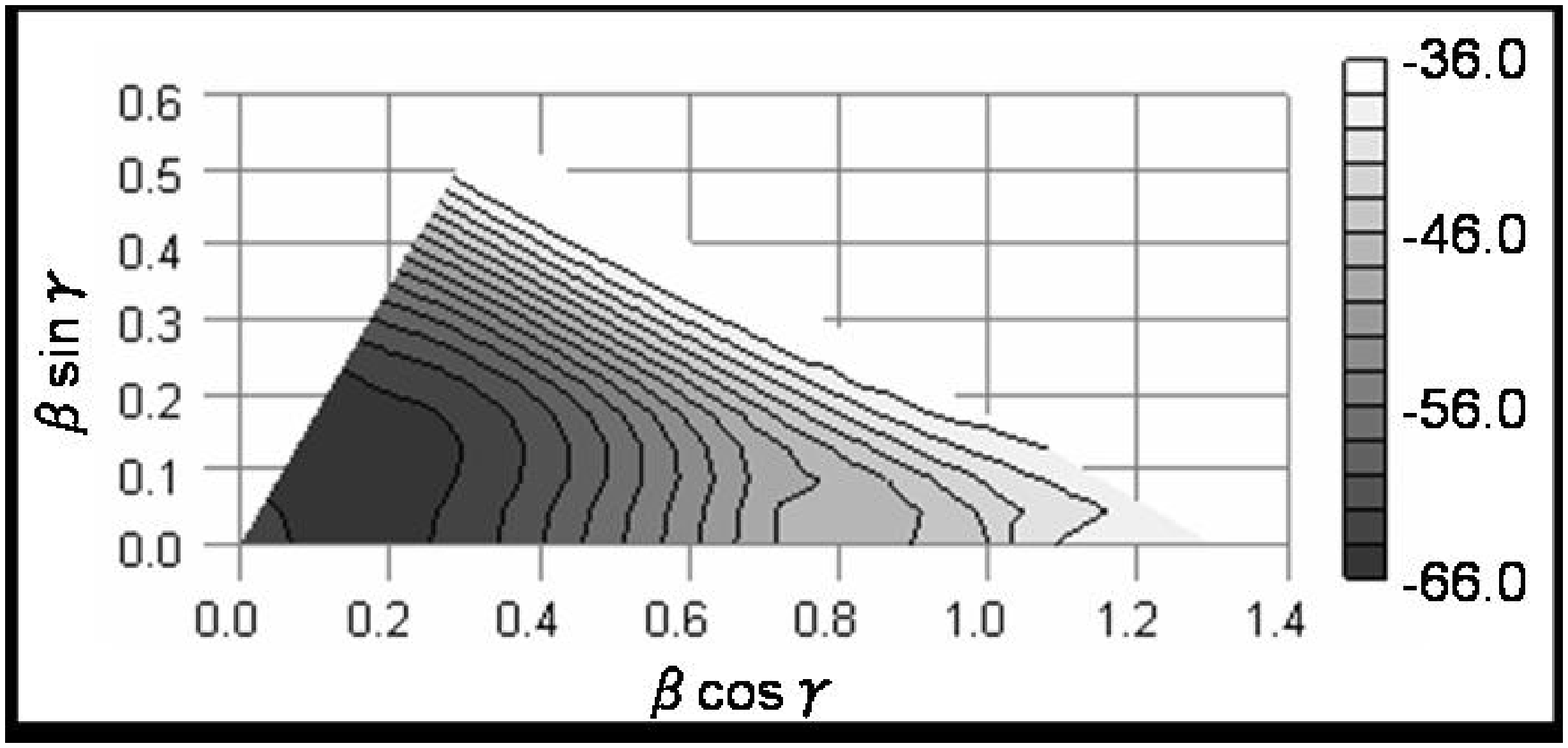}}
	\hfill
	\parbox{\halftext}{\includegraphics[width=6.6cm, bb=10 12 693 323, clip]{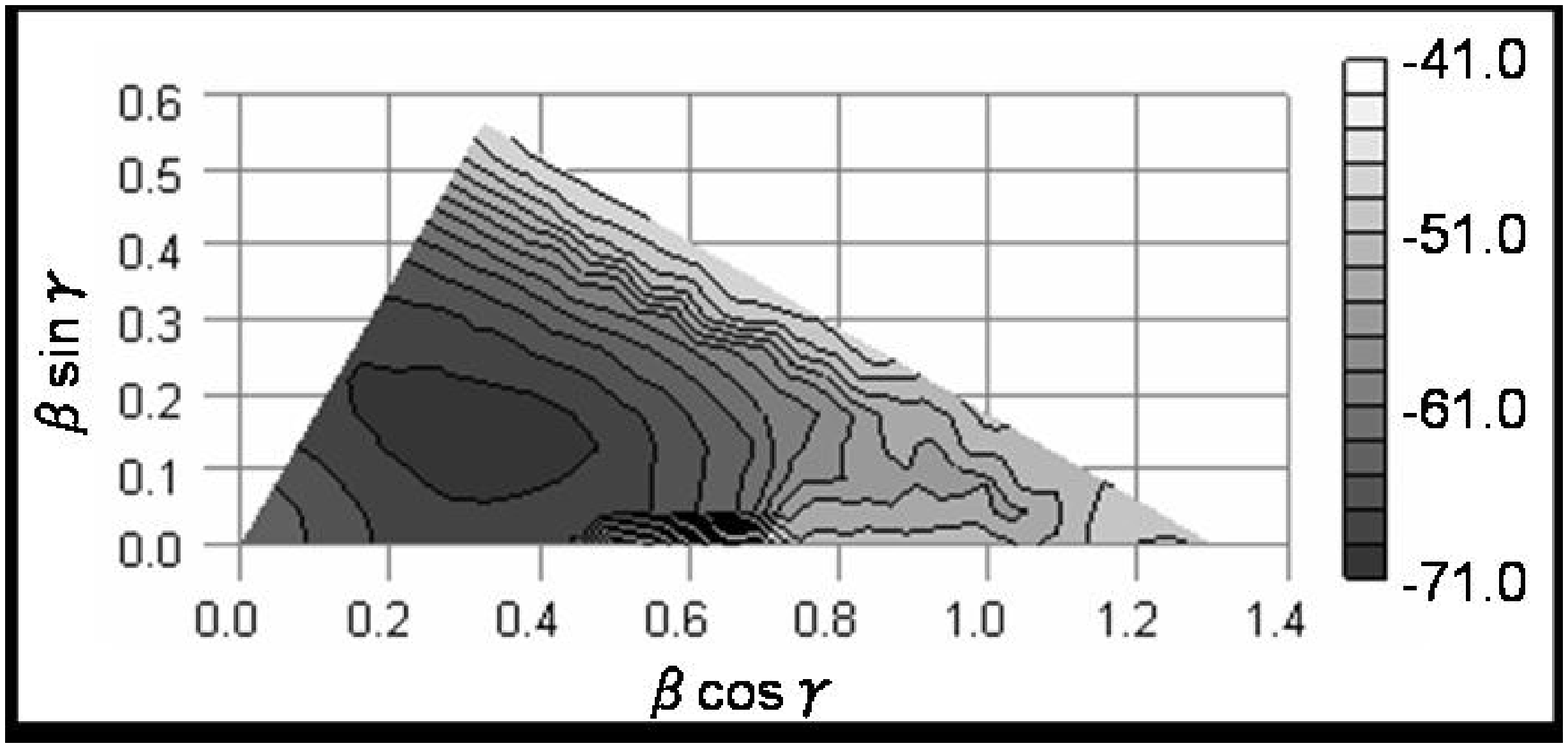}}
	\caption{Energy surfaces of $^{11}$B on the $\beta$-$\gamma$ plane.
	The left shows the energy for the negative-parity states and 
	the right shows that for the $3/2^{-}$ states after the total-angular-momentum projection.}
	\label{11B_energy_surface}
\end{figure}

\begin{figure}[tb]
	\centering
	\includegraphics[angle=-90,width=6cm]{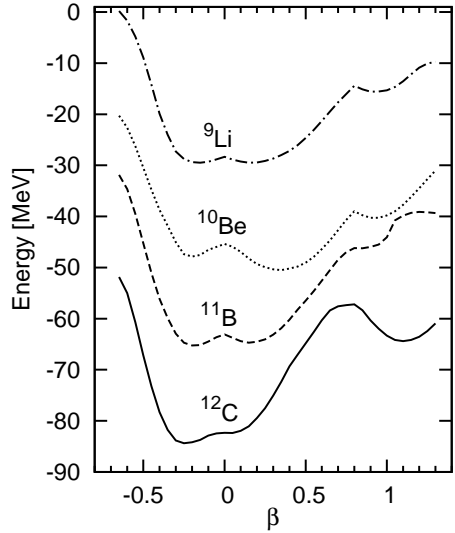}
	\caption{Positive-parity energy curves on the symmetric axes, $\gamma = 0^{\circ}$ and $60^{\circ}$ axes.
	The energy of the prolate states on the $\gamma = 0^{\circ}$ line is displayed 
	in positive $\beta$ and that of the oblate states on the $\gamma = 60^{\circ}$ line 
	is displayed in negative $\beta$.}
	\label{symmetric_energy_surface}
\end{figure}

First, we performed variational calculations with the $\beta$-$\gamma$ constraint
at a total of 196 mesh points of the triangle lattice on the $\beta$-$\gamma$ plane.
We obtained energy surfaces as functions of $\beta$ and $\gamma$.
The calculated energy surfaces on the $\beta$-$\gamma$ plane
before and after the total-angular-momentum projection for $^{10}$Be,
$^{12}$C, $^9$Li, and $^{11}$B  are shown 
in Figs.~\ref{10Be_energy_surface},~\ref{12C_energy_surface},~\ref{9Li_energy_surface}, and~\ref{11B_energy_surface}, respectively. 
We also show the energy curves along the axial symmetric line, 
$\gamma=0^\circ$ and $\gamma=60^\circ$, in Fig.~\ref{symmetric_energy_surface}.

In Fig.~\ref{10Be_energy_surface} for $^{10}$Be,
the left panel shows the energy of the positive-parity states before the total-angular-momentum projection,
and the right panel shows the results 
for $0^{+}$ states calculated using the total-angular-momentum projection after the variation.
We call the former the positive-parity energy surface and the latter 
the $0^{+}$ energy surface.
In the positive-parity energy surface, the minimum and a local minimum 
exist in the prolate region along the $\gamma=0^{\circ}$ line.
The minimum point is at $(\beta \cos \gamma, \beta \sin \gamma) = (0.35, 0.0)$
in the normal deformation region,
while the local minimum exists 
around $(\beta \cos \gamma, \beta \sin \gamma) = (0.9, 0.0)$
in the large deformation region.
After the total-angular-momentum projection, the shape of the energy surface changes 
because the energy gain due to the projection tends to be large
in the large $\beta$ region and triaxial region compared with that in the 
spherical or axial symmetric regions.
As a result, the minimum point of the $0^{+}$ energy surface
shifts to $(\beta \cos \gamma, \beta \sin \gamma) = (0.55, 0.09)$ 
in the finite $\gamma$ region. 
The local minimum point of the $0^+$ energy surface also
has a finite $\gamma$ value as $(\beta \cos \gamma, \beta \sin \gamma) = (1.03, 0.04)$. 
Around the local minimum, the energy surface is rather flat. 
As we show later, the largely deformed excited band 
appears from this region in the GCM energy spectra.

The calculated energy surfaces of $^{12}$C are shown in 
Fig.~\ref{12C_energy_surface}.
The positive-parity energy surface and the $0^{+}$ energy surface are displayed in the 
left and right panels, respectively.
The minimum point of the positive-parity energy surface is at
$(\beta \cos \gamma, \beta \sin \gamma) = (0.13, 0.22)$, which indicates an oblate deformation. 
In contrast to $^{10}$Be, which energetically favors the prolate deformation for any $\beta$, 
in the case of $^{12}$C, the prolate deformation around 
$(\beta \cos \gamma, \beta \sin \gamma) = (0.8, 0.0)$ is unfavored.
There is a peak with 7 MeV height at this point, and a valley 
goes from the oblate energy minimum to the prolate region keeping away from this peak.
After the total-angular-momentum projection, the minimum point of the $0^{+}$ energy surface becomes
$(\beta \cos \gamma, \beta \sin \gamma) = (0.35, 0.17)$. This indicates that
the deformation of the energy minimum state changes from the oblate shape to the 
triaxial shape after the total-angular-momentum projection.
There is also the flat region around $(\beta \cos \gamma, \beta \sin \gamma) = (1.0, 0.1)$
in the $0^+$ energy surface.
In the region around $(\beta \cos \gamma, \beta \sin \gamma) = (0.6, 0.0)$, 
the $0^{+}$ energy surface has a large discontinuity 
because the intrinsic states in this region contain very small $0^{+}$ components.

The energy surfaces of $^{9}$Li are shown in Fig.~\ref{9Li_energy_surface}.
The left and right panels show the negative-parity energy surface before
the total-angular-momentum projection
and the $3/2^{-}$ energy surface after the total-angular-momentum projection, respectively.
In the negative-parity energy surface, it is found that 
the energy minimum state has a small deformation.
A local minimum exists around $(\beta \cos \gamma, \beta \sin \gamma) = (1.0, 0.0)$
in the large prolate region.
After the total-angular-momentum projection, 
the minimum point of the $3/2^{-}$ energy surface becomes
$(\beta \cos \gamma, \beta \sin \gamma) = (0.38, 0.13)$.
In the region near $(\beta \cos \gamma, \beta \sin \gamma) = (1.03, 0.04)$, 
the energy surface is rather flat, and there exists a local minimum.
It is interesting that 
the behavior of the $3/2^{-}$ energy surface of $^{9}$Li is 
qualitatively similar to that of the $0^+$ energy surface of $^{10}$Be.

The results of $^{11}$B are shown in Fig.~\ref{11B_energy_surface}.
The negative-parity energy surface in the left panel shows 
the energy minimum point at $(\beta \cos \gamma, \beta \sin \gamma) = (0.13, 0.13)$.
After the total-angular-momentum projection, the minimum point of the $3/2^{-}$ energy surface 
becomes $(\beta \cos \gamma, \beta \sin \gamma) = (0.33, 0.13)$ as seen in the right panel of
Fig.~\ref{11B_energy_surface}. In the large prolate region, 
a valley is found around $(\beta \cos \gamma, \beta \sin \gamma) = (0.9, 0.1)$
in a similar way to $^{12}$C.

\subsection{Structures on the $\beta$-$\gamma$ plane}

In this section, we discuss the intrinsic structures obtained 
using the $\beta$-$\gamma$ constraint AMD, while paying attention to the cluster aspect.

We analyze the spatial configurations of the Gaussian centers $\{\bm{Z}_1,\bm{Z}_2,\cdots,\bm{Z}_A\}$ 
and proton and neutron density distributions of each intrinsic wave functions $|\Phi(\beta,\gamma) \rangle$.
We also investigate the difference 
$\Tilde{\rho}_{n} - \Tilde{\rho}_{p}$ between the neutron density $\Tilde{\rho}_{n}$ and
proton density $\Tilde{\rho}_{p}$ for neutron-rich nuclei to observe valence neutron behaviors.
Here, the density $\Tilde{\rho}$ is that integrated along the $y$-axis as
\begin{align}
	& \Tilde{\rho} (x, z) \equiv \int dy \rho (\bm{r}), \\ 
	& \rho (\bm{r}) \equiv \langle \Phi (\beta, \gamma) | \sum_{i} \delta (\bm{r} - \bm{r}_{i}) |\Phi (\beta, \gamma) \rangle.
	\label{density_distribution}
\end{align}

\begin{figure}[tb]
	\centering
	{\tabcolsep=0.7mm
	\begin{tabular}{cccccc}
	\vspace{0.3cm}
	($\beta \cos \gamma$, $\beta \sin \gamma$) &
	\hspace{-0.2cm} $\Tilde{\rho}_{p}$ & \hspace{-0.2cm} $\Tilde{\rho}_{n}$ 
	& \hspace{-0.15cm}  $\Tilde{\rho}_{n} - \Tilde{\rho}_{p}$ & \hspace{-0.15cm}  $\Tilde{\rho}_{n}^{\prime} - \Tilde{\rho}_{p}^{\prime}$ \\
	\vspace{-2.2cm}
	\begin{minipage}{.17\linewidth}\vspace{-1.7cm}(a) ($0.35$, $0.00$)\end{minipage} &
	\includegraphics[width=1.8cm]{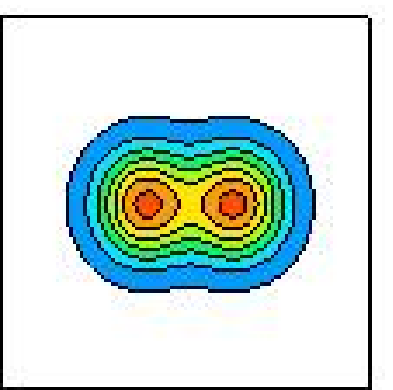} &
	\includegraphics[width=1.8cm]{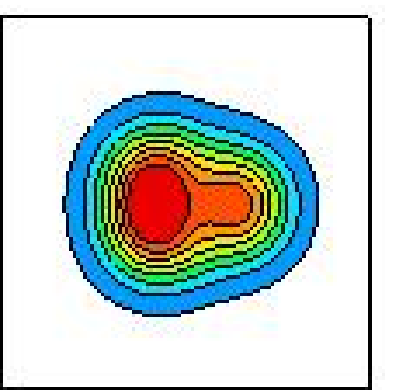} &
	\includegraphics[width=1.8cm]{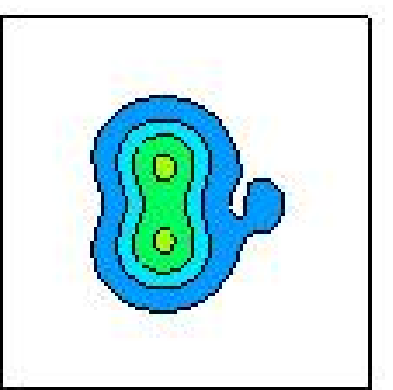} & 
	\includegraphics[width=1.8cm]{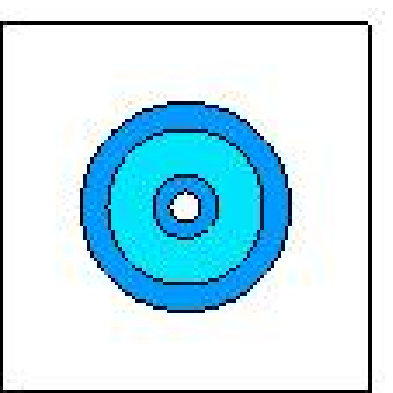} &
	\begin{minipage}{.10\linewidth}\vspace{-3.2cm}[1/fm$^{2}$]\end{minipage} \\
	\begin{minipage}{.17\linewidth}\vspace{-1.7cm}(b) ($0.55$, $0.09$)\end{minipage} &
	\includegraphics[width=1.8cm]{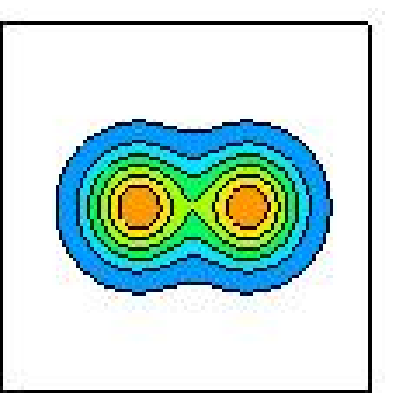} &
	\includegraphics[width=1.8cm]{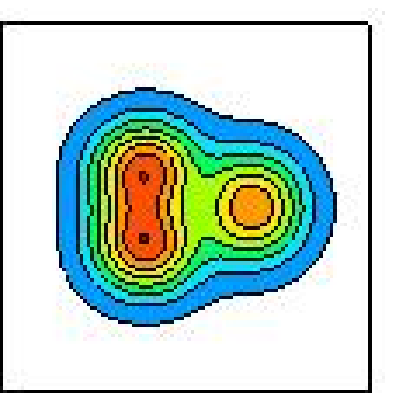} &
	\includegraphics[width=1.8cm]{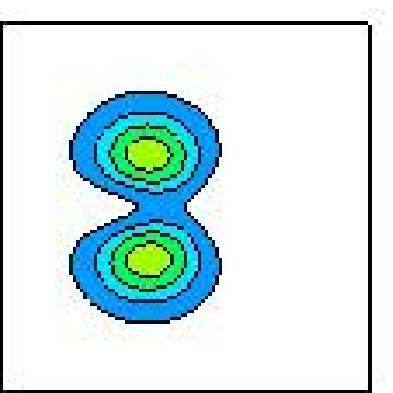} &
	\includegraphics[width=1.8cm]{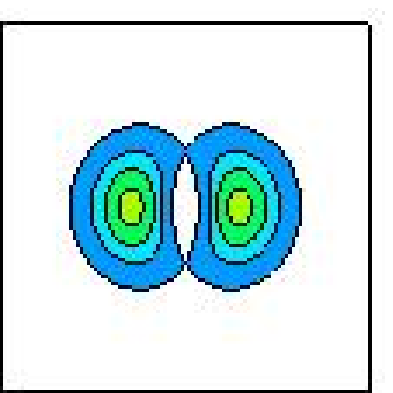} &
	\includegraphics[width=1.5cm,bb = 0 0 60 160,clip]{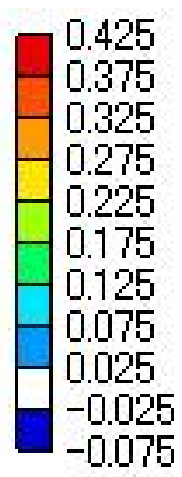} \\
	\begin{minipage}{.17\linewidth}\vspace{-1.7cm}(c) ($0.28$, $0.39$)\end{minipage} &
	\includegraphics[width=1.8cm]{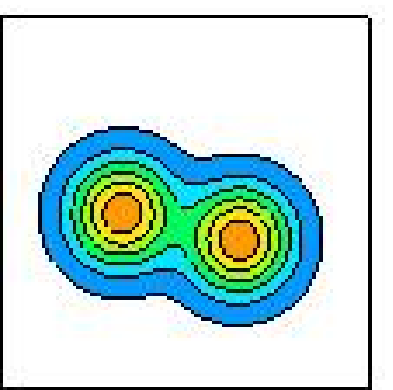} &
	\includegraphics[width=1.8cm]{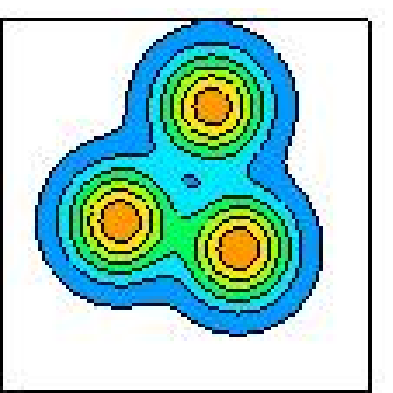} &
	\includegraphics[width=1.8cm]{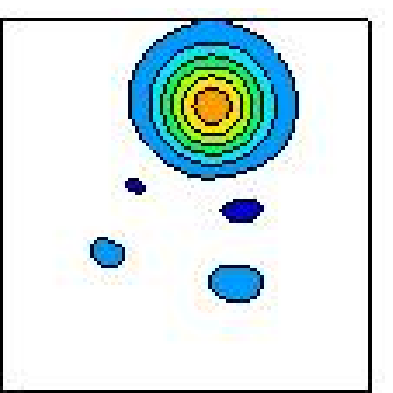} \\
	\begin{minipage}{.17\linewidth}\vspace{-1.7cm}(d) ($1.03$, $0.04$)\end{minipage} &
	\includegraphics[width=1.8cm]{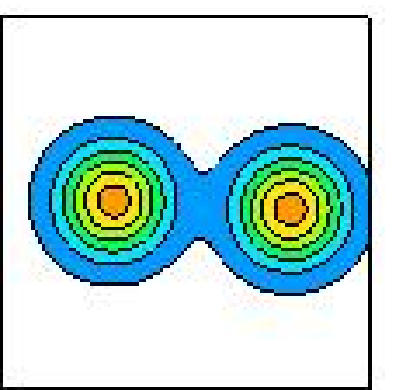} &
	\includegraphics[width=1.8cm]{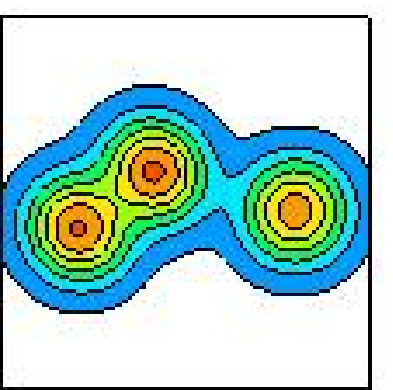} &
	\includegraphics[width=1.8cm]{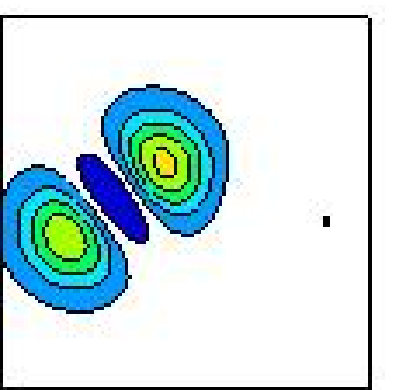} \\
	\begin{minipage}{.17\linewidth}\vspace{-1.7cm}(e) ($0.70$, $0.26$)\end{minipage} &
	\includegraphics[width=1.8cm]{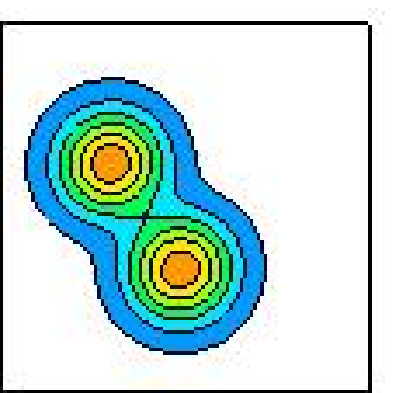} &
	\includegraphics[width=1.8cm]{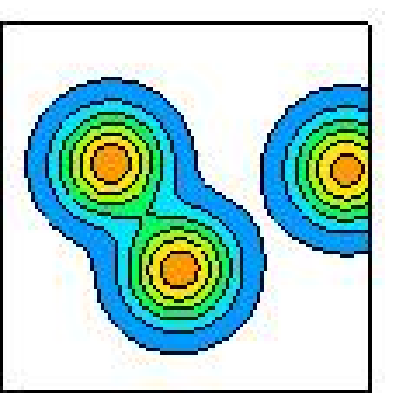} &
	\includegraphics[width=1.8cm]{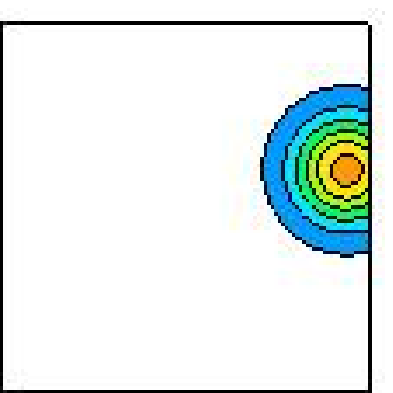} \\
	\end{tabular}}
	\caption{Density distributions of $^{10}$Be.
	The proton density $\Tilde{\rho}_{p}$, neutron density $\Tilde{\rho}_{n}$, 
	difference between the neutron and proton densities 
	$\Tilde{\rho}_{n} - \Tilde{\rho}_{p}$ integrated along the $y$-axis
	are illustrated in the left three panels.
	The density distributions of the intrinsic wave functions at 
	(a) $(\beta \cos \gamma,\beta \sin \gamma)=(0.35,0.00)$,
	(b) $(\beta \cos \gamma,\beta \sin \gamma)=(0.55,0.09)$,
	(c) $(\beta \cos \gamma,\beta \sin \gamma)=(0.28,0.39)$,
	(d) $(\beta \cos \gamma,\beta \sin \gamma)=(1.03,0.04)$, and 
	(e) $(\beta \cos \gamma,\beta \sin \gamma)=(0.70,0.26)$ on the $\beta$-$\gamma$ plane are shown. 
	For the states (a) and (b), the difference between neutron and proton densities 
	$\Tilde{\rho}_{n}^{\prime} - \Tilde{\rho}_{p}^{\prime}$, which is integrated 
	along the $z$-axis but not the $y$-axis, is illustrated in the right panels.
	The size of the box is 10 $\times$ 10 fm$^{2}$.}
	\label{density_10Be}
\end{figure}

The density distributions for $^{10}$Be are illustrated in Fig.~\ref{density_10Be}.
Figure~\ref{density_10Be}(a) shows the density distributions for $|\Phi (\beta,\gamma) \rangle$ at
$(\beta \cos \gamma, \beta \sin \gamma) = (0.35, 0.00)$, which is 
the energy minimum point of the positive-parity energy surface.
In this state, two $\alpha$ clusters are formed as seen in the dumbbell shape of the proton density.
On the other hand, excess neutrons are found to occupy the axial symmetric orbitals and regarded 
to be in the $\pi_{3/2}$-like molecular orbitals. 
In fact, the expectation value of the squared neutron spin $\langle S_{n}^{2} \rangle$ is 
1.07 and is close to the value $\langle S_{n}^{2} \rangle=1$ 
for the $(\pi_{3/2})^{2}$ configuration.
Figure~\ref{density_10Be}(b) shows the density distribution of the energy minimum state
with $(\beta \cos \gamma, \beta \sin \gamma)=(0.55, 0.09)$ in the $0^{+}$ energy surface.
It is found that two $\alpha$ clusters develop further.
In this state, the density distribution of excess neutrons is not axial symmetric but
is extended to the $x$-axis direction resulting in the axial asymmetric.
This means that excess neutrons occupy the $p_{x}$-like orbitals to have the finite $\gamma$ value
instead of the axial symmetric molecular orbitals.
Furthermore, $\langle S_{n}^{2} \rangle$ becomes small to be 0.47. 
This indicates that excess neutrons have the larger $S_{n}=0$ component 
than in the Fig.~\ref{density_10Be}(a) state and is consistent with the 
dominant $(p_{x})^{2}$ configuration.
This indicates that, after the total-angular-momentum
projection, the energy minimum state has 
the more developed 2$\alpha$ cluster structure and more two-neutron correlations of a spin-zero pair 
than those before the projection.
Figures~\ref{density_10Be}(c), (d), and (e) are the density distributions for 
typical structures with oblate, prolate, and triaxial deformations in the large $\beta$ region.
Interestingly, two $\alpha$ clusters develop well in all these states 
with the large deformations, but differences are found in the distributions of excess neutrons.
In Fig.~\ref{density_10Be}(c) for the oblate state, 
excess neutrons distribute far from the 2$\alpha$ core indicating a dineutron cluster.
In Fig.~\ref{density_10Be}(d)  for the almost prolate state, 
excess neutrons distribute around one of the two $\alpha$ cluster causing the 
$^{6}$He correlation.
In terms of the molecular 
orbital model, this state is also regarded as the molecular structure with 
the excess neutrons in the $\sigma_{1/2}$-like orbitals as discussed in 
Ref.~\citen{En'yo_10Be_99}.
Figure~\ref{density_10Be}(e) for the triaxial state shows 
two $\alpha$ clusters and a dineutron cluster with a configuration other than that in the oblate state.
When we pay attention to the neutron densities of these three states, 
it is found that the neutron structure changes 
from the triangular configuration in the oblate region (c) to the narrow chainlike structure in the
prolate region (d) as $\gamma$ becomes small.
Therefore, the structures of $^{10}$Be as functions of deformation parameters,
$\beta$ and $\gamma$, are understood as follows. 
As $\beta$ increases on the $\gamma = 0^{\circ}$ axis, 
a cluster structure with two $\alpha$ clusters and surrounding two neutrons develops, 
and the excess neutron orbitals change from the $\pi_{3/2}$-like orbitals 
to the $\sigma_{1/2}$-like orbitals.
On the other hand, as $\gamma$ becomes finite from $0^{\circ}$ to
$60^{\circ}$, two neutrons have more the $S_{n}=0$ correlation.
When the deformation is sufficiently large,
a dineutron cluster develops and drops out of the 2$\alpha$ core.

\begin{figure}[tb]
	\centering
	{\tabcolsep=0.7mm
	\begin{tabular}{cccc}
	\vspace{0.3cm}
	($\beta \cos \gamma$, $\beta \sin \gamma$) &
	\hspace{-0.2cm} $\Tilde{\rho}_{p}$ & \hspace{-0.2cm} $\Tilde{\rho}_{n}$ \\
	\vspace{-2.2cm}
	\begin{minipage}{.17\linewidth}\vspace{-1.7cm}(a) ($0.13$, $0.22$)\end{minipage} &
	\includegraphics[width=1.8cm]{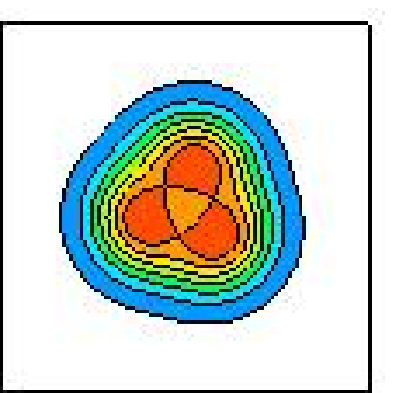} &
	\includegraphics[width=1.8cm]{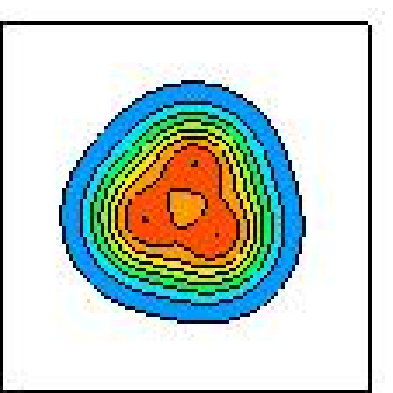} &
	\begin{minipage}{.10\linewidth}\vspace{-3.2cm}[1/fm$^{2}$]\end{minipage} \\
	\begin{minipage}{.17\linewidth}\vspace{-1.7cm}(b) ($0.35$, $0.17$)\end{minipage} &
	\includegraphics[width=1.8cm]{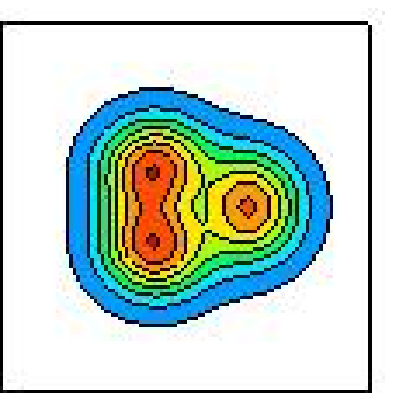} &
	\includegraphics[width=1.8cm]{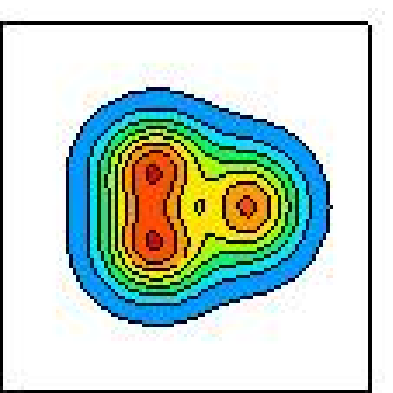} &
	\includegraphics[width=1.5cm,bb = 0 0 60 160,clip]{legend.eps} \\
	\begin{minipage}{.17\linewidth}\vspace{-1.7cm}(c) ($0.33$, $0.48$)\end{minipage} &
	\includegraphics[width=1.8cm]{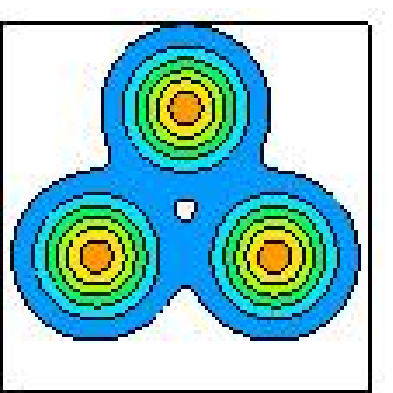} &
	\includegraphics[width=1.8cm]{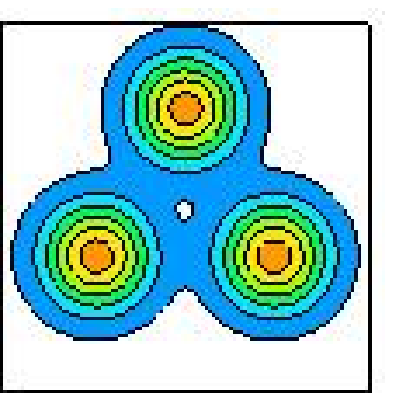} \\
	\begin{minipage}{.17\linewidth}\vspace{-1.7cm}(d) ($1.13$, $0.04$)\end{minipage} &
	\includegraphics[width=1.8cm]{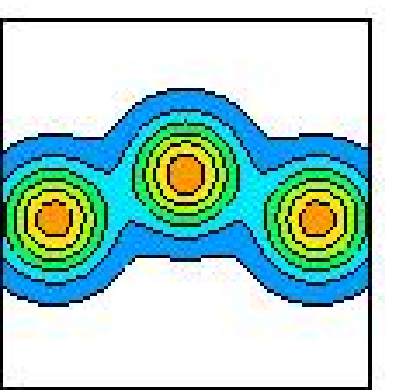} &
	\includegraphics[width=1.8cm]{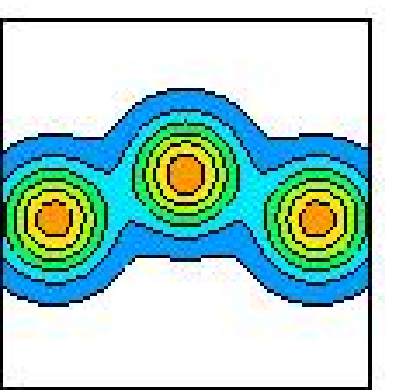} \\
	\begin{minipage}{.17\linewidth}\vspace{-1.7cm}(e) ($0.65$, $0.26$)\end{minipage} &
	\includegraphics[width=1.8cm]{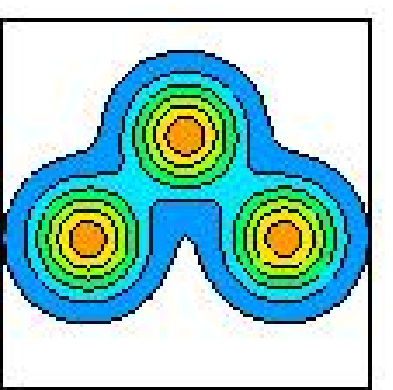} &
	\includegraphics[width=1.8cm]{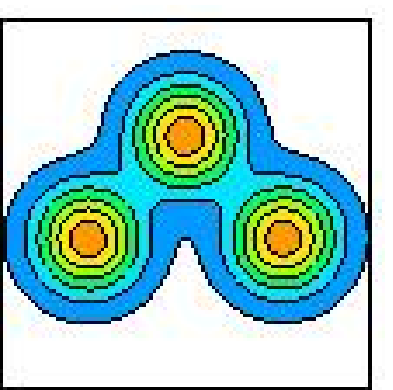} \\
	\end{tabular}}
	\caption{Density distributions of $^{12}$C.
	The proton density $\Tilde{\rho}_{p}$ and neutron density $\Tilde{\rho}_{n}$ 
	are illustrated in the left and right columns, respectively. 
	The density distributions of the intrinsic wave functions at 
	(a) $(\beta \cos \gamma,\beta \sin \gamma)=(0.13,0.22)$,
	(b) $(\beta \cos \gamma,\beta \sin \gamma)=(0.35,0.17)$,
	(c) $(\beta \cos \gamma,\beta \sin \gamma)=(0.33,0.48)$,
	(d) $(\beta \cos \gamma,\beta \sin \gamma)=(1.13,0.04)$, and 
	(e) $(\beta \cos \gamma,\beta \sin \gamma)=(0.65,0.26)$ on the $\beta$-$\gamma$ plane are shown.
	The size of the box is 10 $\times$ 10 fm$^{2}$.}
	\label{density_12C}
\end{figure}

Next, we show the density distributions of $^{12}$C in Fig.~\ref{density_12C}.
In all the obtained wave functions $|\Phi (\beta,\gamma) \rangle$, 
the proton and neutron densities are almost the same with each other.
Figure~\ref{density_12C}(a) is the density distribution for the minimum point 
$(\beta \cos \gamma, \beta \sin \gamma)=(0.13, 0.22)$ in the positive-parity energy surface.
In this wave function, the centers of the single-particle Gaussian wave packets 
gather around the origin and there is no spatially developed cluster structure. 
The expectation value of squared intrinsic spin of protons and that of neutrons are 0.45,
which indicates a large component of non-3$\alpha$ configurations.
In other words, this state is considered to be the shell-model-like structure 
with the dominant $p_{3/2}$ subshell closed component.
Figure~\ref{density_12C}(b) is the density distribution for the 
minimum point $(\beta \cos \gamma, \beta \sin \gamma)=(0.35, 0.17)$ in the $0^{+}$ energy surface.
In this state, an $\alpha$ cluster begins to develop compared with the state (a) 
for the energy minimum before the total-angular-momentum projection.
However, single-particle Gaussian wave packets still gather around the origin, and this state
is regarded as the intermediate between the shell-model-like structure and the cluster structure.
In the large deformation region, three $\alpha$ clusters develop well in $^{12}$C.
As is expected, various configurations of 3 $\alpha$ clusters appear, depending on the deformation
parameters, $\beta$ and $\gamma$.
Figures~\ref{density_12C}(c), (d), and (e) are typical density distributions 
for the oblate, prolate, and triaxial deformed states, respectively.
It is found that the equilateral-triangular configuration,
linear-chainlike structure, and obtuse-angle-triangular configuration arise
in the oblate state (c), prolate state (d), and triaxial state (e), respectively.
Thus, on the $\beta$-$\gamma$ plane for $^{12}$C system, 
we obtained various structures such as 
the shell-model structure in the small $\beta$ region
and the cluster structure in the large $\beta$ region.
In particular, three $\alpha$ clusters develop as the deformation parameter $\beta$ 
becomes large. Various spatial configurations of three $\alpha$ clusters such as 
the linear chain and the equilateral-triangular structures are obtained 
as a function of the triaxiality $\gamma$.

\begin{figure}[tb]
	\centering
	{\tabcolsep=0.7mm
	\begin{tabular}{ccccc}
	\vspace{0.3cm}
	($\beta \cos \gamma$, $\beta \sin \gamma$) &
	\hspace{-0.2cm} $\Tilde{\rho}_{p}$ & \hspace{-0.2cm} $\Tilde{\rho}_{n}$ & \hspace{-0.15cm}  $\Tilde{\rho}_{n} - \Tilde{\rho}_{p}$ \\
	\vspace{-2.2cm}
	\begin{minipage}{.17\linewidth}\vspace{-1.7cm}(a) ($0.15$, $0.00$)\end{minipage} &
	\includegraphics[width=1.8cm]{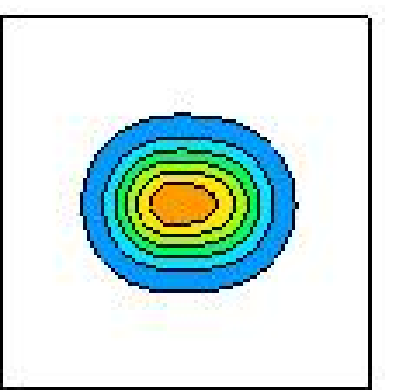} &
	\includegraphics[width=1.8cm]{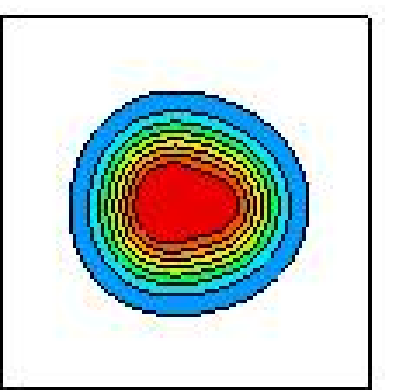} &
	\includegraphics[width=1.8cm]{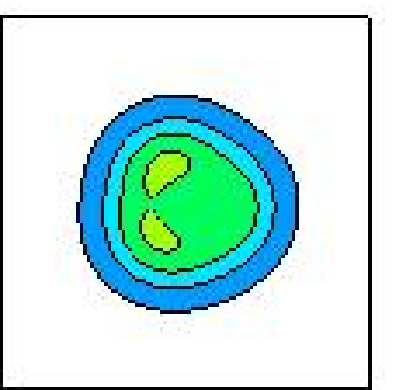} &
	\begin{minipage}{.10\linewidth}\vspace{-3.2cm}[1/fm$^{2}$]\end{minipage} \\
	\begin{minipage}{.17\linewidth}\vspace{-1.7cm}(b) ($0.38$, $0.13$)\end{minipage} &
	\includegraphics[width=1.8cm]{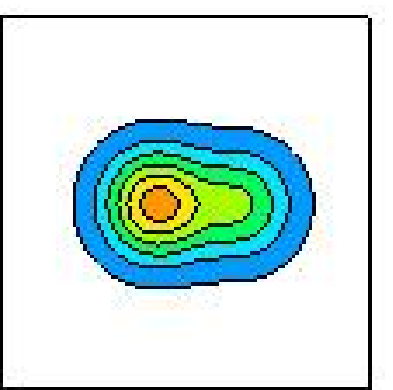} &
	\includegraphics[width=1.8cm]{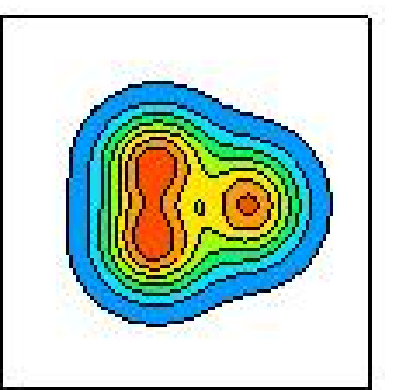} &
	\includegraphics[width=1.8cm]{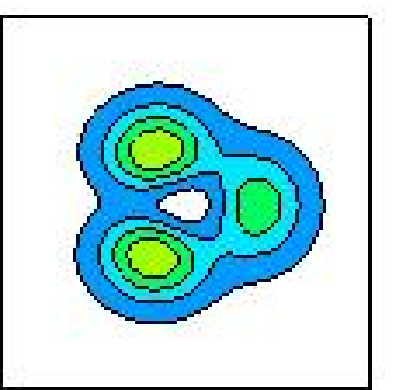} &
	\includegraphics[width=1.5cm,bb = 0 0 60 160,clip]{legend.eps} \\
	\begin{minipage}{.17\linewidth}\vspace{-1.7cm}(c) ($0.28$, $0.39$)\end{minipage} &
	\includegraphics[width=1.8cm]{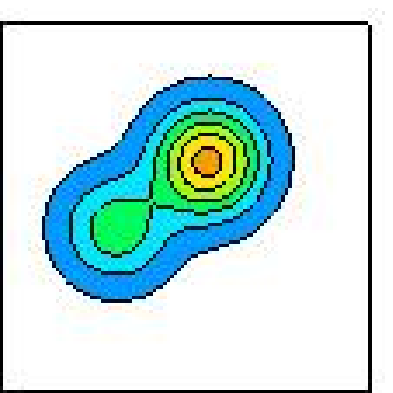} &
	\includegraphics[width=1.8cm]{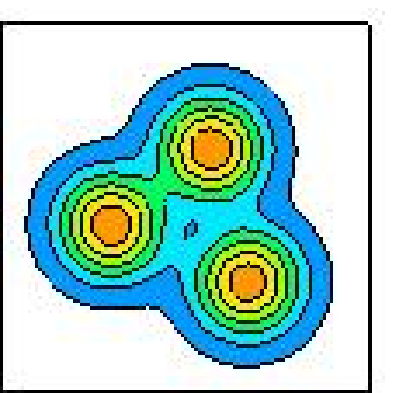} &
	\includegraphics[width=1.8cm]{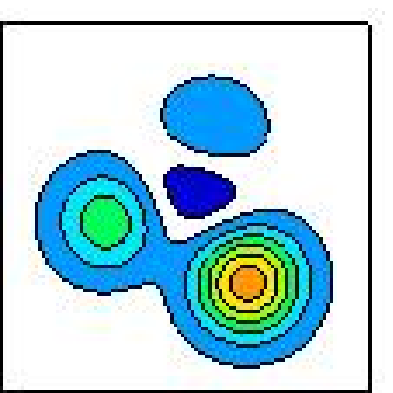} \\
	\begin{minipage}{.17\linewidth}\vspace{-1.7cm}(d) ($1.03$, $0.04$)\end{minipage} &
	\includegraphics[width=1.8cm]{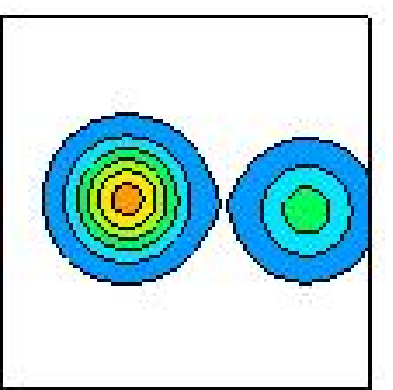} &
	\includegraphics[width=1.8cm]{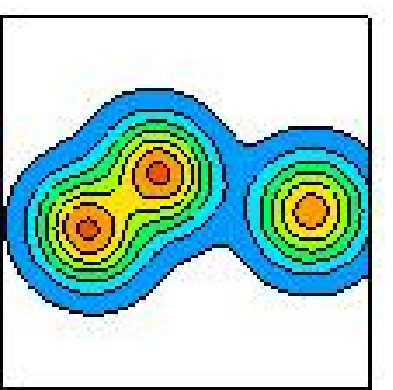} &
	\includegraphics[width=1.8cm]{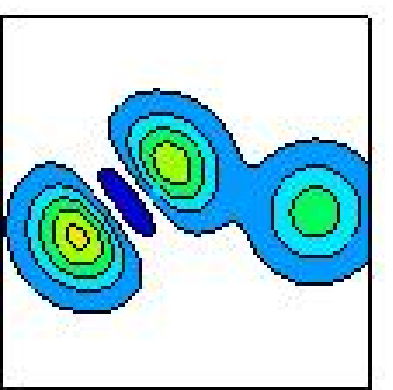} \\
	\begin{minipage}{.17\linewidth}\vspace{-1.7cm}(e) ($0.70$, $0.26$)\end{minipage} &
	\includegraphics[width=1.8cm]{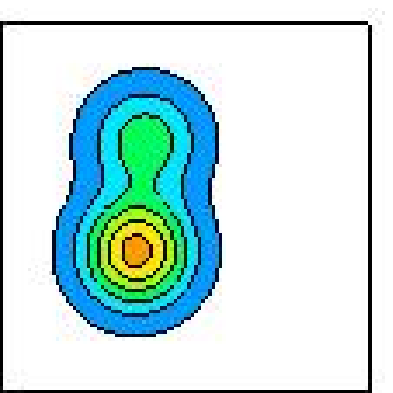} &
	\includegraphics[width=1.8cm]{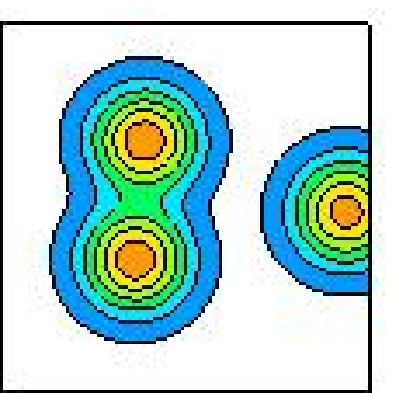} &
	\includegraphics[width=1.8cm]{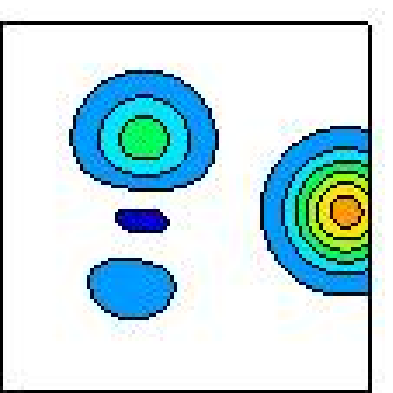} \\
	\end{tabular}}
	\caption{Density distributions of $^{9}$Li.
	The proton density $\Tilde{\rho}_{p}$, neutron density $\Tilde{\rho}_{n}$, 
	and difference between the neutron and proton densities $\Tilde{\rho}_{n} - \Tilde{\rho}_{p}$
	are illustrated in the left, middle, and right columns, respectively. 
	The density distributions of the intrinsic wave functions at 
	(a) $(\beta \cos \gamma,\beta \sin \gamma)=(0.15,0.00)$,
	(b) $(\beta \cos \gamma,\beta \sin \gamma)=(0.38,0.13)$,
	(c) $(\beta \cos \gamma,\beta \sin \gamma)=(0.28,0.39)$,
	(d) $(\beta \cos \gamma,\beta \sin \gamma)=(1.03,0.04)$, and 
	(e) $(\beta \cos \gamma,\beta \sin \gamma)=(0.70,0.26)$ on the $\beta$-$\gamma$ plane are shown.
	The size of the box is 10 $\times$ 10 fm$^{2}$.}
	\label{density_9Li}
\end{figure}

Figure~\ref{density_9Li} shows the density distributions of $^{9}$Li.
Figures~\ref{density_9Li}(a) and (b) correspond to the 
energy minimum states before and after the total-angular-momentum
projection, respectively.
Figures~\ref{density_9Li}(c), (d), and (e) are density distributions of typical structures
in the oblate, prolate, and triaxial states with large deformations, respectively.
The cluster features in $^{9}$Li are a good analogy to those in $^{10}$Be, and
therefore, they can be understood in a similar way as the above discussion for 
$^{10}$Be by replacing one $\alpha$ cluster in the $^{10}$Be system to a triton cluster in $^9$Li.
Namely, as the deformation parameter $\beta$ increases on the $\gamma = 0^{\circ}$ axis, 
the structure changes from the spherical shell-model state 
to the developed cluster structure with the $\alpha+t$ cluster and
surrounding two excess neutrons in the axial symmetric molecular orbitals.
As the triaxiality $\gamma$ increases from 0$^\circ$ to
60$^\circ$, a pair of two neutrons drops out of the $\alpha+t$ cluster 
and describes the two-neutron correlation.

\begin{figure}[tb]
	\centering
	{\tabcolsep=0.7mm
	\begin{tabular}{ccccc}
	\vspace{0.3cm}
	($\beta \cos \gamma$, $\beta \sin \gamma$) &
	\hspace{-0.2cm} $\Tilde{\rho}_{p}$ & \hspace{-0.2cm} $\Tilde{\rho}_{n}$ & \hspace{-0.15cm}  $\Tilde{\rho}_{n} - \Tilde{\rho}_{p}$ \\
	\vspace{-2.2cm}
	\begin{minipage}{.17\linewidth}\vspace{-1.7cm}(a) ($0.13$, $0.13$)\end{minipage} &
	\includegraphics[width=1.8cm]{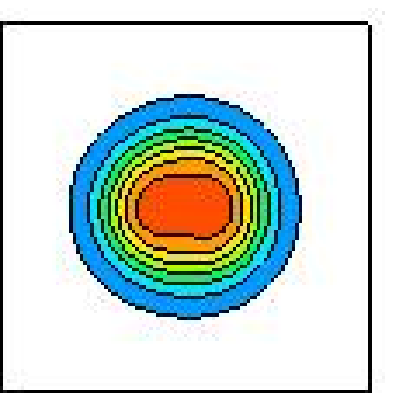} &
	\includegraphics[width=1.8cm]{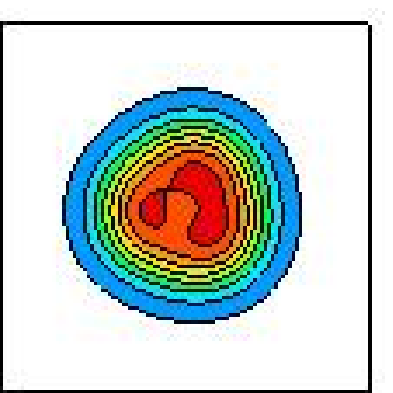} &
	\includegraphics[width=1.8cm]{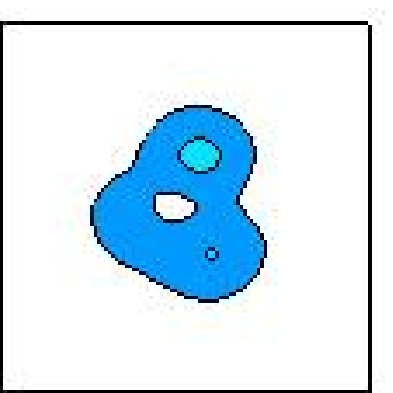} &
	\begin{minipage}{.10\linewidth}\vspace{-3.2cm}[1/fm$^{2}$]\end{minipage} \\
	\begin{minipage}{.17\linewidth}\vspace{-1.7cm}(b) ($0.33$, $0.13$)\end{minipage} &
	\includegraphics[width=1.8cm]{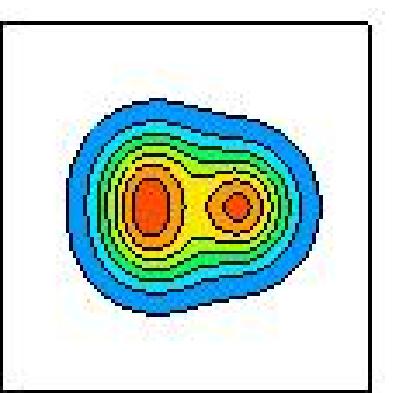} &
	\includegraphics[width=1.8cm]{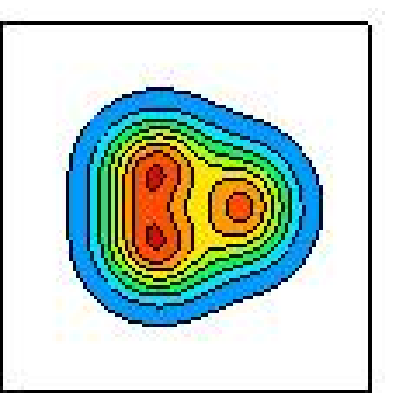} &
	\includegraphics[width=1.8cm]{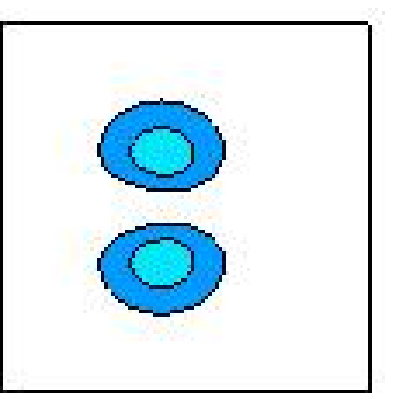} &
	\includegraphics[width=1.5cm,bb = 0 0 60 160,clip]{legend.eps} \\
	\begin{minipage}{.17\linewidth}\vspace{-1.7cm}(c) ($0.33$, $0.48$)\end{minipage} &
	\includegraphics[width=1.8cm]{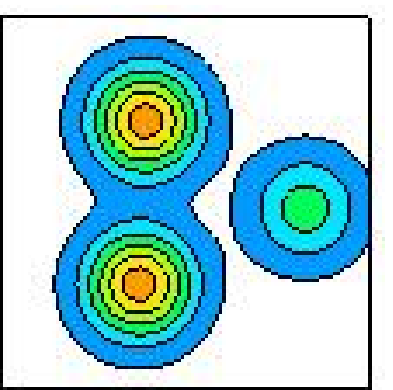} &
	\includegraphics[width=1.8cm]{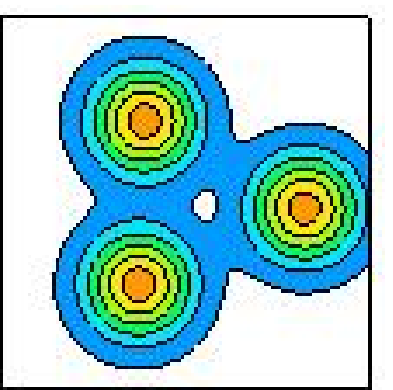} &
	\includegraphics[width=1.8cm]{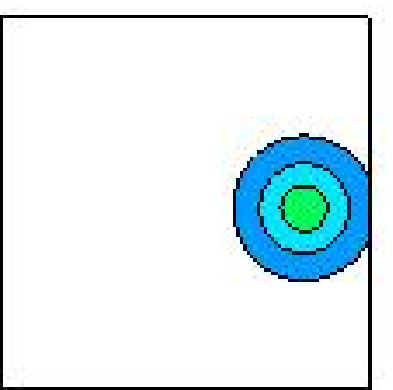} \\
	\begin{minipage}{.17\linewidth}\vspace{-1.7cm}(d) ($1.13$, $0.04$)\end{minipage} &
	\includegraphics[width=1.8cm]{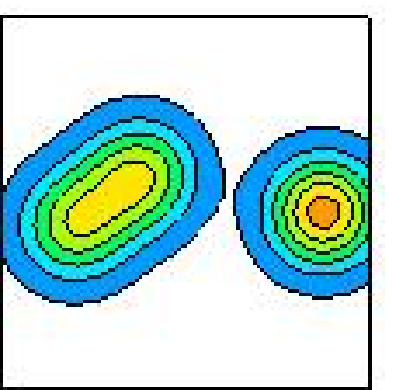} &
	\includegraphics[width=1.8cm]{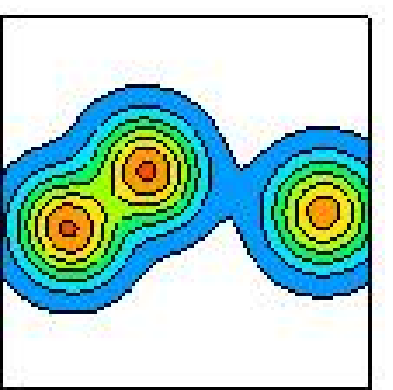} &
	\includegraphics[width=1.8cm]{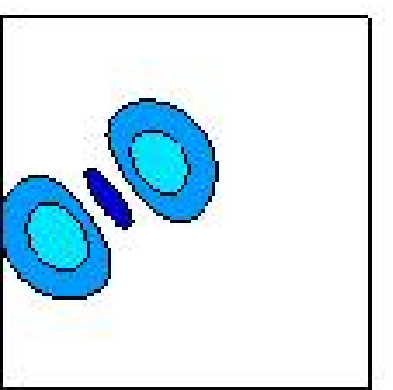} \\
	\begin{minipage}{.17\linewidth}\vspace{-1.7cm}(e) ($0.65$, $0.26$)\end{minipage} &
	\includegraphics[width=1.8cm]{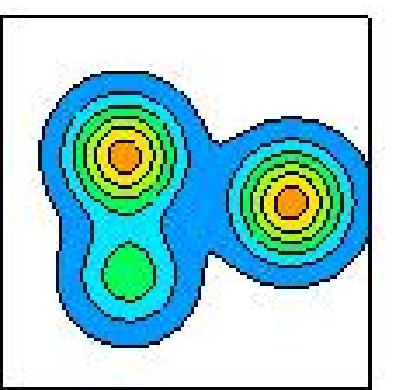} &
	\includegraphics[width=1.8cm]{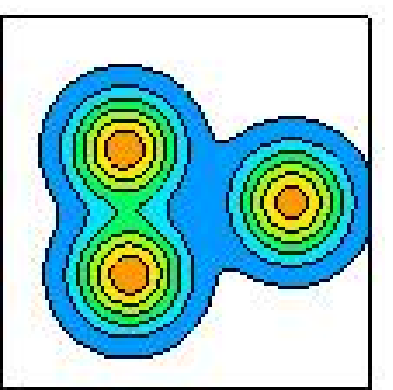} &
	\includegraphics[width=1.8cm]{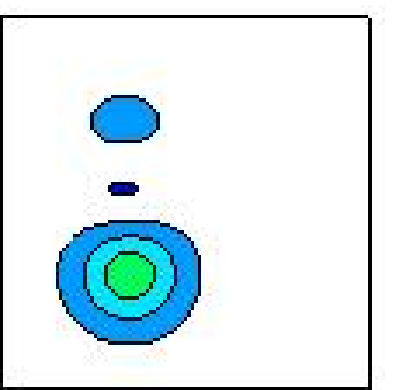} \\
	\end{tabular}}
	\caption{Density distributions of $^{11}$B.
	The proton density $\Tilde{\rho}_{p}$, neutron density $\Tilde{\rho}_{n}$, 
	and difference between the neutron and proton densities $\Tilde{\rho}_{n} - \Tilde{\rho}_{p}$
	are illustrated in the left, middle, and right columns, respectively. 
	The density distributions of the intrinsic wave functions at 
	(a) $(\beta \cos \gamma,\beta \sin \gamma)=(0.13,0.13)$,
	(b) $(\beta \cos \gamma,\beta \sin \gamma)=(0.33,0.13)$,
	(c) $(\beta \cos \gamma,\beta \sin \gamma)=(0.33,0.48)$,
	(d) $(\beta \cos \gamma,\beta \sin \gamma)=(1.13,0.04)$, and 
	(e) $(\beta \cos \gamma,\beta \sin \gamma)=(0.65,0.26)$ on the $\beta$-$\gamma$ plane are shown.
	The size of the box is 10 $\times$ 10 fm$^{2}$.}
	\label{density_11B}
\end{figure}

The density distributions of $^{11}$B are illustrated in Fig.~\ref{density_11B}.
The shell and cluster features of $^{11}$B have a good correspondence to those of $^{12}$C.
In the small deformation region, the neutron density has an almost spherical shape 
and is regarded as the dominant $p_{3/2}$-shell closed configuration.
As the deformation parameter $\beta$ increases, the structure changes from a shell-model state to 
a developed cluster structure via the shell-cluster competition region.
The cluster features in the large deformation region are understood
with the $2\alpha+t$ clustering in $^{11}$B instead of the 3$\alpha$ cluster structure in $^{12}$C. 
Similarly to the case of $^{12}$C, as a function of $\gamma$, 
we obtained various types of spatial configurations of three-center clusters such as 
the equilateral-triangular structure in the oblate deformation 
and the linear-chain structure in the prolate deformation.

We summarize the shell and cluster features of these $N=6$ systems 
obtained by the present method of the $\beta$-$\gamma$ constraint AMD on the $\beta$-$\gamma$ plane.
In the small $\beta$ region, $\beta < 0.2 \sim 0.3$, shell-model-like structures appear.
In particular, the neutron structure has the dominant $p_{3/2}$-shell closed
configurations with the spherical shape.
In this region, the centers of the single-particle Gaussian wave packets 
gather around the origin to keep the deformation small and 
there are no spatially developed cluster structures. 
In the large $\beta$ region, cluster structures develop well.
Depending on the $\gamma$ parameter, various cluster configurations appear.
That is, in the prolate region near the $\gamma = 0^{\circ}$ line,
two-body cluster structures or linear-chainlike structures 
develop well as $\beta$ becomes large.
In the oblate region near the $\gamma = 60^{\circ}$ line, 
three-body cluster structures with equilateral-triangular or isosceles-triangular configurations appear.
In the triaxial region, various triangle configurations of the three-body clusters are obtained.

\subsection{GCM results}

We superposed the wave functions obtained using the $\beta$-$\gamma$ constraint AMD with the GCM to calculate 
the energy spectra of $^{10}$Be and $^{12}$C.

\begin{figure}[tb]
\centering
	\includegraphics[angle=-90,width=11.2cm]{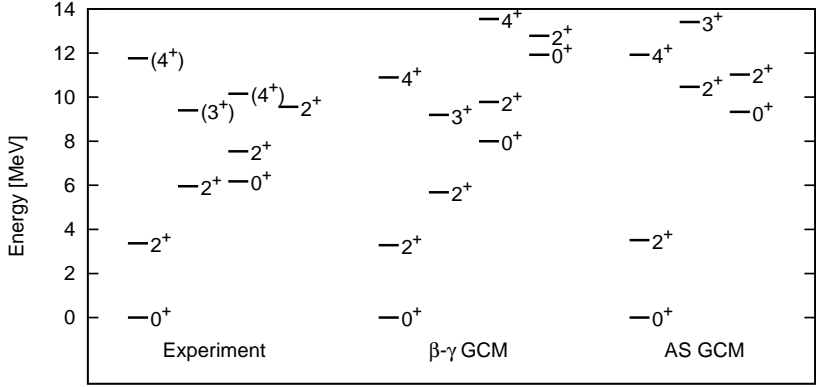}
	\caption{Energy levels of the positive-parity states in $^{10}$Be.
		Four columns on the left are the experimental data, four columns in the middle are the two-dimensional $\beta$-$\gamma$ GCM results,
		and three columns on the right are the axial symmetric GCM results.
		In the two-dimensional $\beta$-$\gamma$ GCM results, we classified the states into four groups.
		The first column is the ground band, 
		the second column is the $K^{\pi}=2^{+}$ side band of the ground band, and
		the third column is the $0^{+}_{2}$ band.
		Other states are plotted in the fourth column.
		The details are described in the text.}
\label{Energy_level_0Be_+}
\end{figure}

First, we describe the results for $^{10}$Be.
The calculated binding energy (B.E.) is $59.2$ MeV, while the experimental one is 64.98 MeV.
In the results, we obtained many excited states above the ground state.
We show the calculated energy levels of $^{10}$Be in Fig.~\ref{Energy_level_0Be_+}
as well as the experimental levels.\cite{Tilley_A=8-9-10_04,Freer_10Be_06,Curtis_Be_06,Bohlen_10Be_07} \ 

The calculated results reproduce well the experimental energy levels.
We classified the calculated states in four groups by analyzing 
the components of the basis wave functions and intrinsic structures.

The calculated $0^{+}_{1}$, $2^{+}_{1}$, and $4^{+}_{1}$ states constitute the ground band.
They have the large component of the basis wave function at $(\beta \cos \gamma, \beta \sin \gamma) = (0.55, 0.09)$.
For instance, the overlap of the $0^{+}$ state with this  
wave function is 87\%.
As discussed before, in this dominant basis wave function, 
two $\alpha$ clusters develop well and excess neutrons occupy $p_{x}$-like orbitals.
The $0^{+}_{1}$ state also has a large overlap with the wave function at $(\beta \cos \gamma, \beta \sin \gamma) = (0.35, 0.00)$
with 70\%. As already mentioned earlier, the 
wave function at $(\beta \cos \gamma, \beta \sin \gamma) = (0.35, 0.00)$ shows the structure with
the $2\alpha$ core and excess neutrons in the $\pi_{3/2}$-like molecular orbitals.
Therefore, roughly speaking, the ground bands can be interpreted as the molecular orbital structure with the $2\alpha$ core
and excess neutrons. However, it should be pointed out that the dominant basis wave function is not the axial symmetric one 
but the triaxially deformed state with 
the excess neutrons in the $p_{x}$-like orbitals. This means that the two-neutron correlation is contained 
significantly in the ground band.

The $2^{+}_{2}$ and $3^{+}_{1}$ states construct the $K^{\pi}=2^{+}$ side band built on the ground band.
These states appear because of the triaxiality of the ground band. 
Namely, the $K^{\pi}=2^{+}$ band arises because the excess neutrons occupy the $p_{x}$-like orbitals
due to the two-neutron correlation.
This is consistent with the theoretical work by Itagaki et al. where the triaxiality in $^{10}$Be was discussed.\cite{Itagaki_10Be_02} \ 

The $0^{+}_{2}$, $2^{+}_{3}$, and $4^{+}_{2}$ states belong to the excited $K^{\pi}=0^{+}$ band
with a large prolate deformation.
In fact, they have a large overlap with the wave function at $(\beta \cos \gamma, \beta \sin \gamma) = (1.03, 0.04)$,
for example, as 64\% overlap in the $0^{+}_{2}$ state.
As we discussed Fig.~\ref{density_10Be}(d), this dominant wave function has the extremely large prolate deformation
due to the developed two $\alpha$ clusters and excess neutrons surrounding the 2$\alpha$ core.

We also obtained other excited states below 14 MeV plotted in the rightmost column in Fig.~\ref{Energy_level_0Be_+}.
The $0^{+}_3$ state around 12 MeV is dominated by the basis wave function with developed two $\alpha$ and dineutron clusters.
This might relate with the $2\alpha$+dineutron condensate state suggested by Itagaki et al.\cite{Itagaki_alpha_condense_07} \ 

The calculated $E2$ transition strengths are listed in Table~\ref{E2_Radius_10Be}~(A).
The calculated strengths, $B(E2, 2^{+}_{1} \rightarrow 0^{+}_{1})=9.4$ $e^{2}$fm$^{4}$ and 
$B(E2, 2^{+}_{1} \rightarrow 0^{+}_{2})=1.2$ $e^{2}$fm$^{4}$, 
agree well with the experimental data, $B(E2, 2^{+}_{1} \rightarrow 0^{+}_{1})=10.24 \pm 0.97$ $e^{2}$fm$^{4}$ and 
$B(E2, 2^{+}_{1} \rightarrow 0^{+}_{2})=0.64 \pm 0.23$ $e^{2}$fm$^{4}$,\cite{Tilley_A=8-9-10_04} \ respectively.
The interband transition strengths between the ground band and its $K^{\pi}=2^{+}$ side band are
$B(E2, 2^{+}_{2} \rightarrow 0^{+}_{1})=0.7$ $e^{2}$fm$^{4}$ and $B(E2, 2^{+}_{2} \rightarrow 2^{+}_{2})=4.2$ $e^{2}$fm$^{4}$.
The origin of these interband transition strengths is 
the recoil effect of the valence neutrons related to two $\alpha$ clusters as discussed in Ref.~\citen{{Itagaki_10Be_02}}.

The calculated root-mean-square radii for mass distributions are listed in Table~\ref{E2_Radius_10Be}~(B).
The calculated radius of the $0^{+}_{1}$ state, 2.39 fm, is close to the 
experimental one, $2.30 \pm 0.02$ fm.\cite{Ozawa_rmsRadius_01} \ 

\begin{table}[tb]
	\caption{Electromagnetic transition strengths $B(E2)$ and radii of $^{10}$Be. 
	(A)  Calculated $B(E2)$ values in the unit of $e^{2}$fm$^{4}$.
	The values in the $\beta$-$\gamma$ GCM column are the two-dimensional $\beta$-$\gamma$ GCM results and 
	the values in the AS GCM column are the axial symmetric GCM results.
	(B) Root-mean-square radii for mass distributions of $^{10}$Be calculated by the 
	two-dimensional $\beta$-$\gamma$ GCM. The unit is fm.}
	\label{E2_Radius_10Be}
	\begin{minipage}[t]{.55\textwidth}
		\centering
		(A) $B(E2)$
		\begin{tabular}{cccc} \hline \hline
		Transitions & $\beta$-$\gamma$ GCM & AS GCM & Experiment \\ \hline
		$2^{+}_{1} \rightarrow 0^{+}_{1}$ & 9.4 & 7.8 & $10.24 \pm 0.97$ \\
		$2^{+}_{1} \rightarrow 0^{+}_{2}$ & 1.2 & 0.1 & $0.64 \pm 0.23$ \\
		$2^{+}_{2} \rightarrow 0^{+}_{1}$ & 0.7 & 0.1 & \\
		$2^{+}_{2} \rightarrow 2^{+}_{1}$ & 4.2 & 0.8 & \\
		\hline \hline
		\end{tabular}
	\end{minipage}
	\hfill
	\begin{minipage}[t]{.40\textwidth}
		\centering
		(B) Root-mean-square radii
		\begin{tabular}{ccc} \hline \hline
		States & $\beta$-$\gamma$ GCM & Experiment \\ \hline
		$0^{+}_{1}$ & 2.39 & $2.30 \pm 0.02$ \\
		$0^{+}_{2}$ & 2.98 & \\
		$0^{+}_{3}$ & 2.96 & \\ 
		\hline \hline
		\end{tabular}
	\end{minipage}
\end{table}

The present results of $^{10}$Be are quantitatively similar to the earlier work studied with 
the molecular orbital model\cite{Itagaki_10Be_00} \ and also to that with AMD.\cite{En'yo_10Be_99} \ 
It indicates that the $\beta$-$\gamma$ constraint AMD is a useful method of preparing
proper basis wave functions that are sufficient to describe the ground and excited states of $^{10}$Be.

To check the advantages of the present method of the two-dimensional constraint
and to discuss the importance of the triaxiality, 
we also calculated the one-dimensional axial symmetric GCM in which the basis is limited to axial symmetric ones
and compared the results with the two-dimensional GCM ones.
This is a GCM calculation performed in a subset of two-dimensional $\beta$-$\gamma$ GCM. 
The number of superposed basis is only 40 on the $\gamma=0^{\circ}$ and $\gamma=60^{\circ}$ axes,
whereas 196 basis are superposed in the two-dimensional $\beta$-$\gamma$ GCM calculation.
The calculated energy levels are shown in Fig.~\ref{Energy_level_0Be_+}.
The calculated B.E. is $58.6$ MeV, which is just $0.6$ MeV smaller than the $\beta$-$\gamma$ GCM result.
We found that axial symmetric basis wave functions are appropriate to reproduce the ground state.
On the other hand, the axial symmetric GCM fails to reproduce some excited states.
For example, the $2^{+}_{2}$ and $3^{+}_{1}$ states are 5.5 MeV higher than the $\beta$-$\gamma$ GCM result.
Furthermore, the agreement of the calculated $B(E2, 2^{+}_{1} \rightarrow 0^{+}_{1})$ with the experimental result
is worse than the $\beta$-$\gamma$ GCM result, as shown in Table~\ref{E2_Radius_10Be}.
These failures are caused by the insufficiency of the GCM basis, which miss the triaxial ones.
As a result of the lack of triaxial wave functions,
interband transition strengths, $B(E2, 2^{+}_{2} \rightarrow 0^{+}_{1})$ and 
$B(E2, 2^{+}_{2} \rightarrow 2^{+}_{2})$, become small.

\begin{figure}[tb]
\centering
	\includegraphics[angle=-90,width=11.2cm]{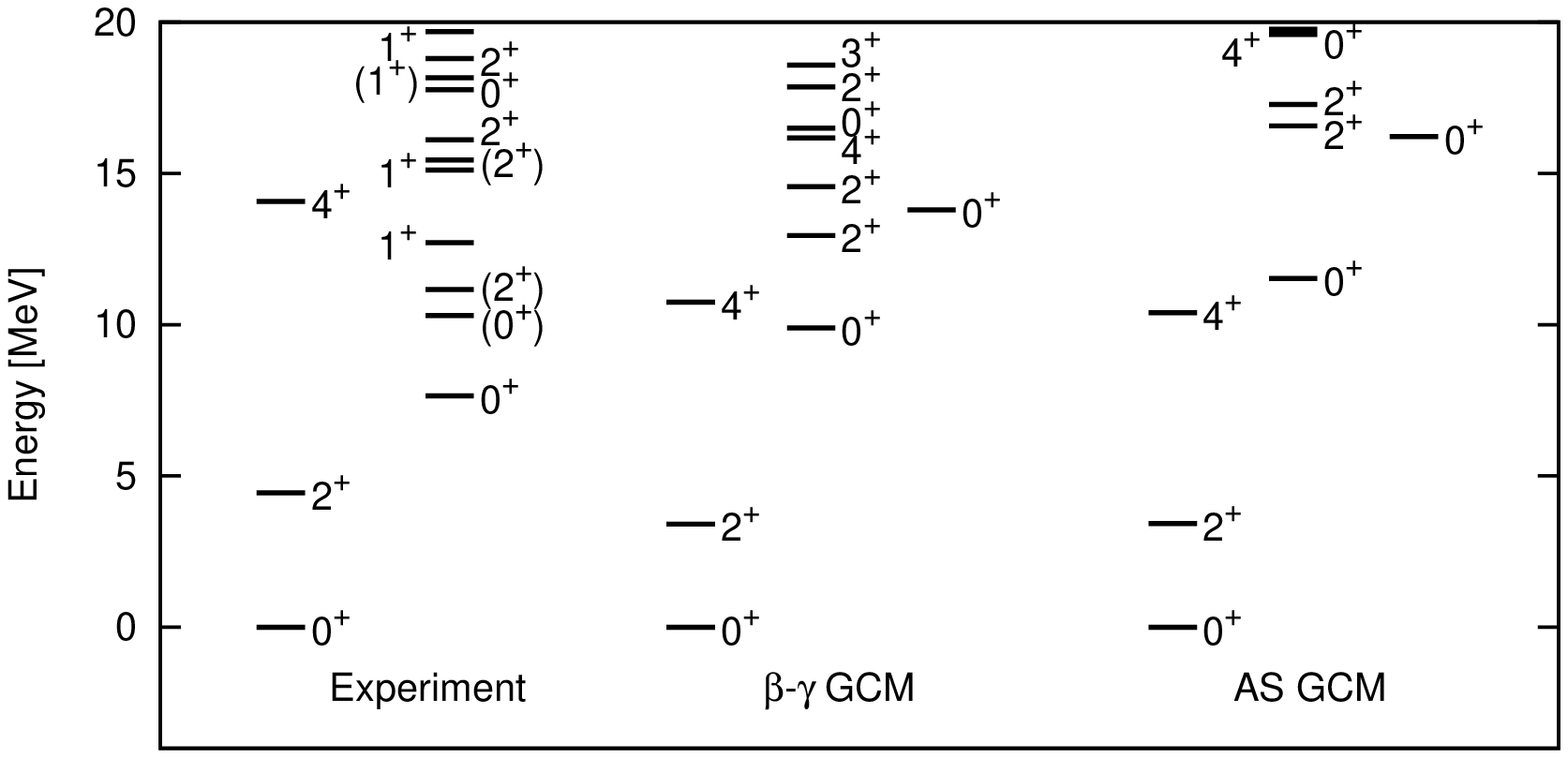}
	\caption{Energy levels of the positive-parity states in $^{12}$C.
		Two columns on the left are the experimental data, three columns in the middle are the two-dimensional $\beta$-$\gamma$ GCM results,
		and three columns on the right are the axial symmetric GCM results.
		In the two-dimensional $\beta$-$\gamma$ GCM results, we classified the states in three groups.
		The first column is the ground band and
		the second column shows the spectra of the states, which have overlap with the basis AMD wave function
		in the broad $\gamma$ area of the large $\beta$ region.
		The third column is the $0^{+}_{3}$ states.
		The details are described in the text.}
\label{Energy_level_12C_+}
\end{figure}

\begin{figure}[tb]
	\parbox{\halftext}{\includegraphics[width=6.6cm, bb=10 12 693 323, clip]{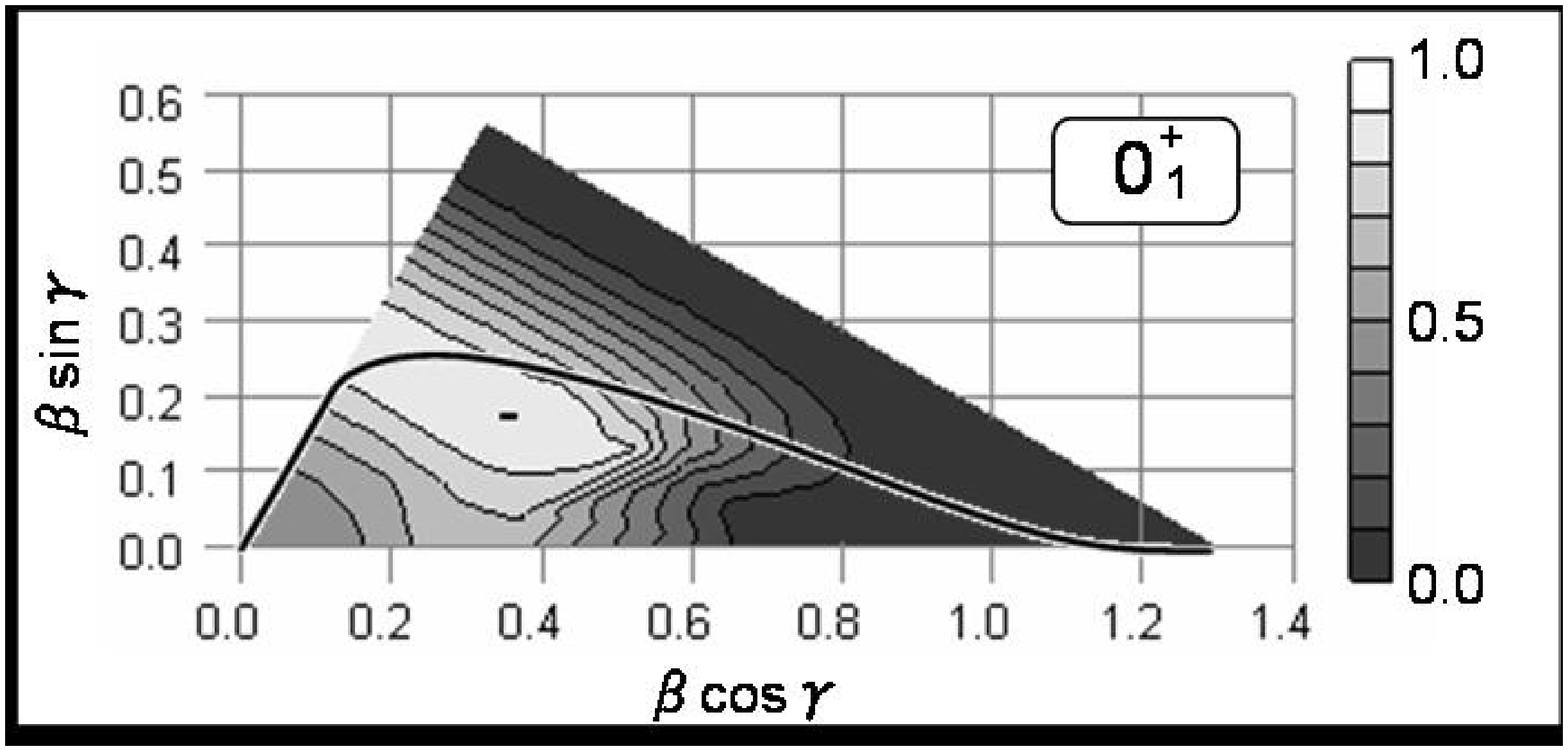}}
	\hfill
	\parbox{\halftext}{\includegraphics[width=6.6cm, bb=10 12 693 323, clip]{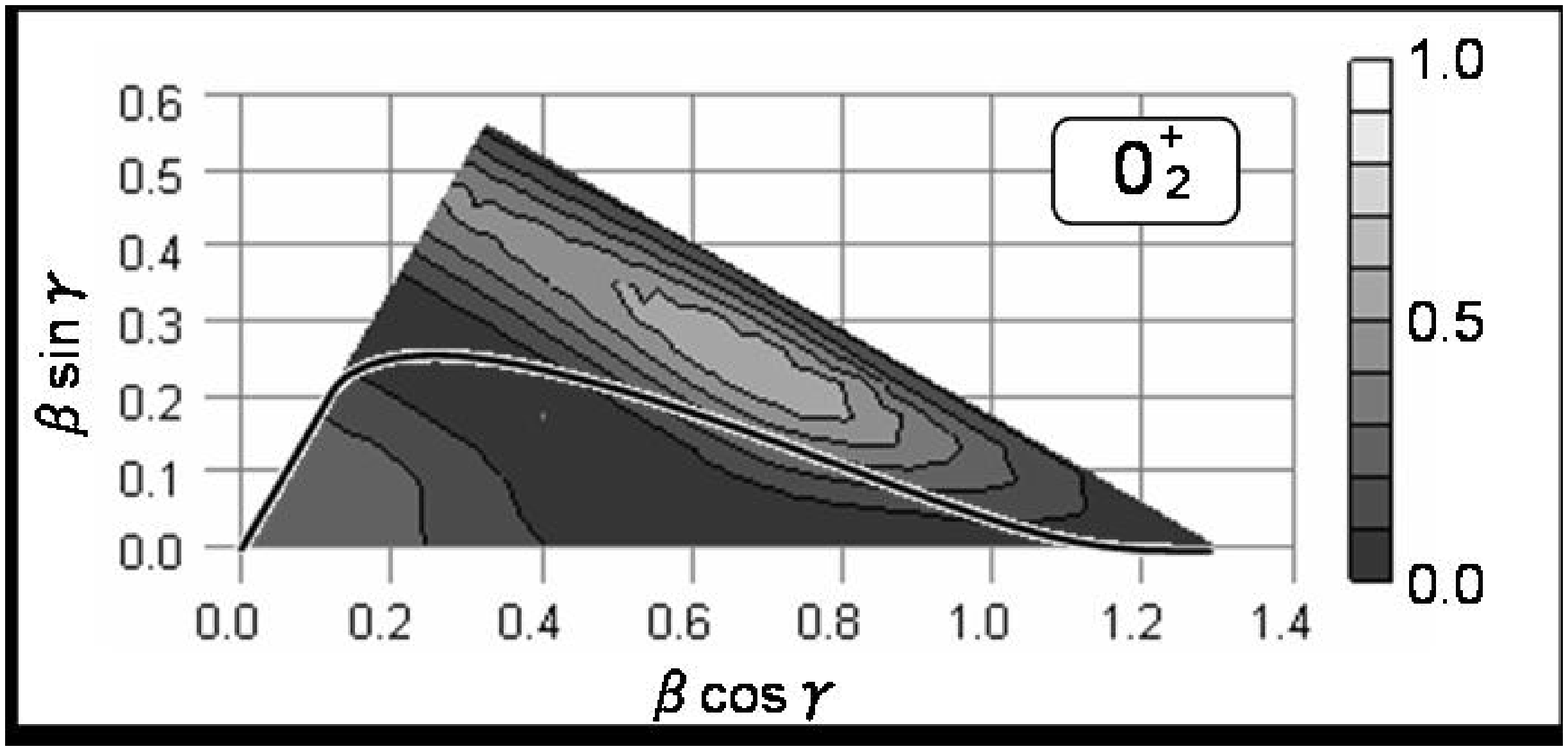}} \\
	\parbox{\halftext}{\includegraphics[width=6.6cm, bb=10 12 693 323, clip]{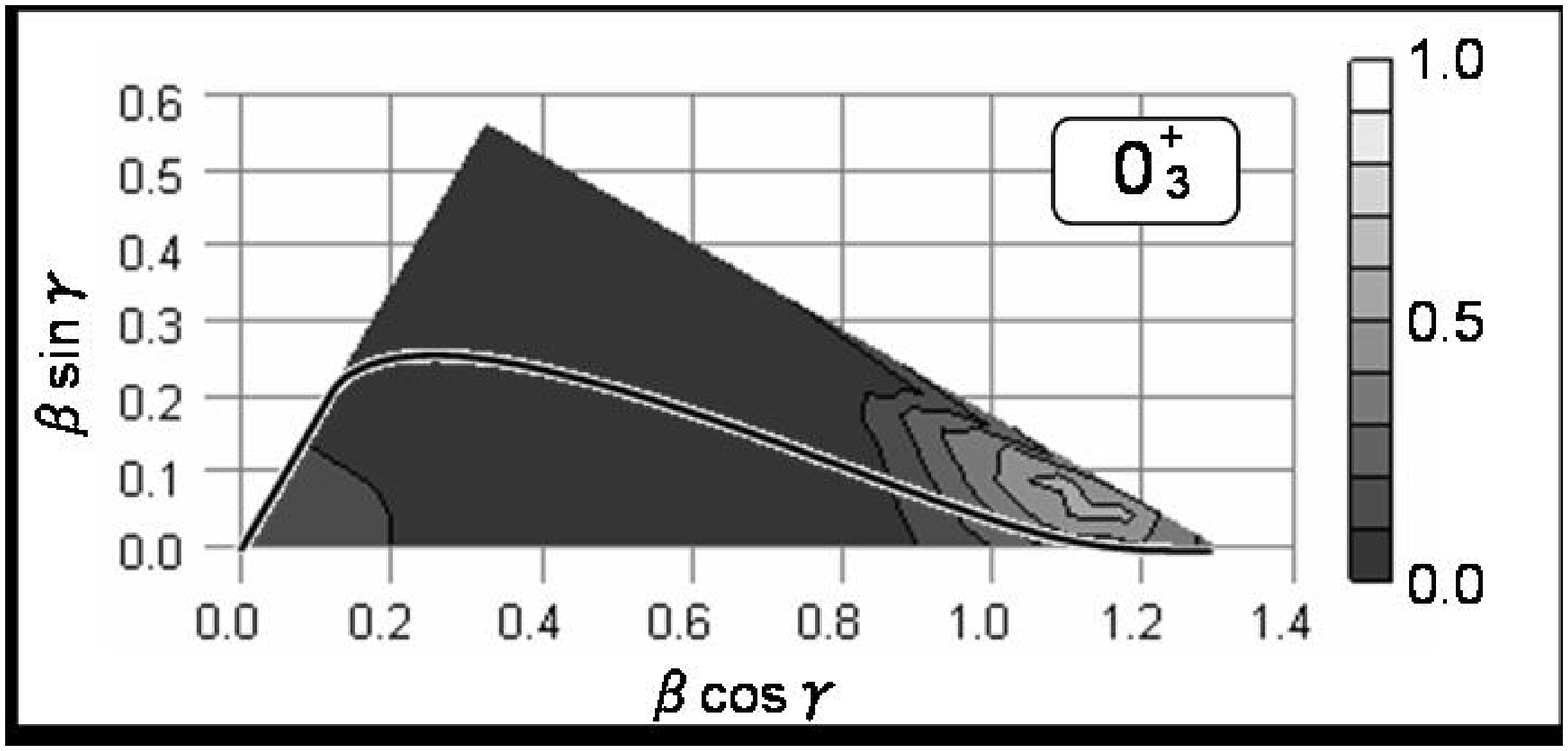}}
	\caption{Overlaps of the GCM wave functions for the $0^{+}$ states of $^{12}$C with 
		each basis AMD wave functions obtained with the $\beta$-$\gamma$ constraint.
		The overlap with a certain ($\beta,\gamma$) point is shown in the contour map.
		The solid lines show the $\beta$ constraint paths, which indicate
		the optimum $\gamma$ of the energy minimum position for a given $\beta$ value. }
\label{overlap_12C_0_+_2}
\end{figure}

Next, we discuss the results of $^{12}$C.
The calculated B.E. is $92.5$ MeV (The experimental B.E. is 92.16 MeV).
In Fig.~\ref{Energy_level_12C_+}, we give the calculated energy levels of $^{12}$C 
as well as the experimental data.\cite{Ajzenber_A=11-12_90} \ 
For the theoretical results, we classified the states in three groups by analyzing overlaps with the basis wave functions.
In Fig.~\ref{overlap_12C_0_+_2}, 
the squared overlaps $|\langle 0^+| P^{J=0}_{00} | \Phi^+ (\beta,\gamma) \rangle|^2$
of the GCM wave functions for the $0^{+}$ states with 
the basis AMD wave functions at each point $(\beta, \gamma)$ are shown.

The $0^{+}_{1}$, $2^{+}_{1}$, and $4^{+}_{1}$ states constitute the ground band. 
They have a large overlap with the basis wave function at $(\beta \cos \gamma, \beta \sin \gamma) = (0.35, 0.17)$,
for instance, as the 90\% overlap in the $0^{+}_{1}$ state.
As discussed earlier, this dominant basis wave function has 
no spatial development of three $\alpha$ clusters but the shell-model-like structure (Fig.~\ref{density_12C}(b)). 
It is interesting that the dominant wave function shows the triaxiality in the intrinsic deformation.

The calculated states displayed in the second column of the right side in Fig.~\ref{Energy_level_12C_+}
have large overlaps with the basis AMD wave functions
in the broad $\gamma$ area of the large $\beta$ region. For example, as shown in Fig.~\ref{overlap_12C_0_+_2}, 
the components of the $0^{+}_{2}$ state do not concentrate on some specific basis wave function but distribute
widely into various basis wave functions. 
In this area of the large $\beta$ region, the basis wave functions describe
various configurations of the developed three $\alpha$ clusters depending on $\beta$ and $\gamma$.
It means that the $0^{+}_{2}$ state is described using the linear combination of various 3$\alpha$-cluster configurations.
This is analogous to the earlier works with
the 3$\alpha$ GCM,\cite{Uegaki_12C_77} \ FMD,\cite{Neff_12C_04} \ and 
AMD calculations,\cite{En'yo_12C_07} \ 
and is also consistent with the suggestion of the $\alpha$ condensate state.\cite{Tohsaki_12C_01,Funaki_12C_03} \ 
This indicates that the present method of $\beta$-$\gamma$ constraint is effective for describing 
the three-body cluster feature of the excited states described by a superposition of multiconfigurations.
In particular, the degree of freedom of the triaxiality $\gamma$ is essential to describe 
this feature. A more detailed discussion will be given in the next section.

The $0^{+}_3$ state has the largest overlap with the wave function at 
$(\beta \cos \gamma, \beta \sin \gamma) = (1.13, 0.04)$ with 54\% overlap.
As shown in Fig.~\ref{density_12C}(d), 
this basis wave function has three $\alpha$ clusters with the linear-chain-like configuration.
This result is consistent with the prediction of the $0^+_3$ state in the 
3$\alpha$ model,\cite{Uegaki_12C_77} \ 
FMD,\cite{Neff_12C_04} \ and AMD calculations.\cite{En'yo_12C_07} \ 

The calculated $E2$ transition strengths are listed in Table~\ref{E2_Radius_12C}~(A).
The calculated strengths, $B(E2, 2^{+}_{1} \rightarrow 0^{+}_{1})=6.0$ $e^{2}$fm$^{4}$ and 
$B(E2, 2^{+}_{1} \rightarrow 0^{+}_{2})=1.9$ $e^{2}$fm$^{4}$, 
agree well with the experimental results, $B(E2, 2^{+}_{1} \rightarrow 0^{+}_{1})=7.59 \pm 0.42$ $e^{2}$fm$^{4}$ and 
$B(E2, 2^{+}_{1} \rightarrow 0^{+}_{2})=2.6 \pm 0.4$ $e^{2}$fm$^{4}$,\cite{Ajzenber_A=11-12_90} \ respectively.

The calculated root-mean-square radii for mass distributions are listed in Table~\ref{E2_Radius_12C}~(B).
The calculated value of the $0^{+}_{1}$ state, 2.31 fm, is close to the experimental value, 
$2.35 \pm 0.02$ fm or $2.31 \pm 0.02$ fm.\cite{Ozawa_rmsRadius_01} \ 

\begin{table}[tb]
	\caption{Electromagnetic transition strengths $B(E2)$ and radii of $^{12}$C. 
	(A)  Calculated $B(E2)$ values in the unit of $e^{2}$fm$^{4}$.
	The values in the $\beta$-$\gamma$ GCM column are the two-dimensional $\beta$-$\gamma$ GCM results and 
	the values in the AS GCM column are the axial symmetric GCM results.
	(B) Root-mean-square radii for mass distributions of $^{12}$C calculated by the 
	two-dimensional $\beta$-$\gamma$ GCM. The unit is fm.}
	\label{E2_Radius_12C}
	\begin{minipage}[t]{.55\textwidth}
		\centering
		(A) $B(E2)$
		\begin{tabular}{cccc} \hline \hline
		Transitions & $\beta$-$\gamma$ GCM & AS GCM & Experiment \\ \hline
		$2^{+}_{1} \rightarrow 0^{+}_{1}$ & 6.0 & 5.3 & $7.59 \pm 0.42$ \\
		$2^{+}_{1} \rightarrow 0^{+}_{2}$ & 1.9 & 1.5 & $2.6 \pm 0.4$ \\ 
		$2^{+}_{2} \rightarrow 0^{+}_{1}$ & 1.1 & 0.4 & \\ 
		$2^{+}_{2} \rightarrow 0^{+}_{2}$ & 58 & 11 & \\
		\hline \hline
		\end{tabular} 
	\end{minipage}
	\hfill
	\begin{minipage}[t]{.40\textwidth}
		\centering
		(B) Root-mean-square radii
		\begin{tabular}{ccc} \hline \hline
		States & $\beta$-$\gamma$ GCM & Experiment \\ \hline
		$0^{+}_{1}$ & 2.31 & $2.35 \pm 0.02$ \\
		& & $2.31 \pm 0.02$ \\
		$0^{+}_{2}$ & 2.90 & \\
		$0^{+}_{3}$ & 3.26 & \\ 
		\hline \hline
		\end{tabular}
	\end{minipage}
\end{table}

The present calculations describe various structures in $^{12}$C.
The shell-model structure appears in the low-lying states, while $3\alpha$-cluster structures are found in
the excited states. 
The shell and cluster coexistence obtained in the present results is consistent with 
the works with the FMD\cite{Neff_12C_04} \ and AMD calculations.\cite{En'yo_12C_07} \ 
The success of the present calculations 
is due to the fact that the $\beta$-$\gamma$ constraint AMD
is effective for obtaining the basis wave functions for the ground and excited states of $^{12}$C.

Let us discuss the energies calculated by the axial symmetric GCM. The energy levels 
obtained by the axial symmetric GCM calculations are shown in Fig.~\ref{Energy_level_12C_+}.
The calculated B.E. is $91.6$ MeV, which is $0.9$ MeV smaller than the $\beta$-$\gamma$ GCM result.
Since the difference from the $\beta$-$\gamma$ GCM result is small, 
the axial symmetric calculation is sufficient to describe the ground state, as in the case of $^{10}$Be.
On the other hand, the level structure of excited states changes considerably.
For example, the $0^{+}_{2}$ and $0^{+}_{3}$ states are 2.6 and 3.4 MeV higher than the $\beta$-$\gamma$ GCM result, respectively.
Additionally, the $B(E2, 2^{+}_{2} \rightarrow 0^{+}_{2})$ value becomes small
as shown in Table~\ref{E2_Radius_12C}.
Clearly, the triaxial basis has an essential role in the description of excited states.

\section{Discussion}\label{discussions}

In calculations of deformed systems with the quadrupole deformation constraint,
one often adopts only one-dimensional $\beta$ constraint
instead of the two-dimensional $\beta$-$\gamma$ constraint.
As mentioned earlier, various structures appear on the $\beta$-$\gamma$ plane and they are
important basis wave functions, especially for the excited states.
In this section, we discuss the advantages of the present method of the two-dimensional constraint
and the importance of the triaxiality described by the $\gamma$ degree of freedom.

The simplest assumption for deformed systems is the axial symmetry 
as carried out in deformed mean-field calculations. In such axial symmetric calculations, 
the deformation parameter $\beta$ is the constraint parameter. 
This corresponds to the $\gamma=0^{\circ}$ and $\gamma=60^{\circ}$ lines on the $\beta$-$\gamma$ plane.
In Fig.~\ref{symmetric_energy_surface}, we have shown the energy curves 
on the $\gamma = 0^{\circ}, 60^{\circ}$ lines for $^9$Li, $^{10}$Be, $^{11}$B, and $^{12}$C.
The energy curves for the axial symmetric deformation show the double-well structure around
$\beta=0$. However, as we mentioned earlier, it does not mean the coexistence of oblate and prolate shapes
but it corresponds to one genuine minimum on the $\beta$-$\gamma$ plane for the ground state.
For example, in the case of $^{10}$Be, 
two local minima at $\beta = -0.2$ and $\beta = 0.35$ are observed in the axial symmetric 
energy curve, as shown in Fig.~\ref{symmetric_energy_surface}.
However, only the $\beta = 0.35$ minimum is the genuine minimum
and the $\beta = -0.2$ minimum comes from the projection of the energy pocket of the 
genuine minimum at $\gamma=0^\circ$ onto the $\gamma=60^\circ$ line.
In the case of $^{12}$C, the local minima in the large prolate deformation around $\beta = 1$ 
is also artificial. This is a part of the valley in the region near 
$(\beta \cos \gamma, \beta \sin \gamma) = (1.0, 0.1)$.
That is, a local minimum on the symmetric axis is sometimes 
an artifact owing to the axial symmetry restriction.
Therefore, one should discuss carefully the energy curve of the axial symmetric 
calculations. In other words, the $\beta$-$\gamma$ constraint method is helpful
for understanding the detailed behavior of the energy surface.

We mention again the one-dimensional axial symmetric GCM.
Needless to say, the basis wave functions with the triaxial deformations are missing in the 
axial symmetric calculations, and therefore, some results for the excited states change considerably 
in comparison with the two-dimensional $\beta$-$\gamma$ GCM calculations.
For example, the $K^{\pi}=2^{+}$ side band of $^{10}$Be 
cannot be described with the axial symmetric GCM calculations as discussed earlier.

\begin{figure}[tb]
	\parbox{\halftext}{\includegraphics[width=6.6cm, bb=10 12 693 323, clip]{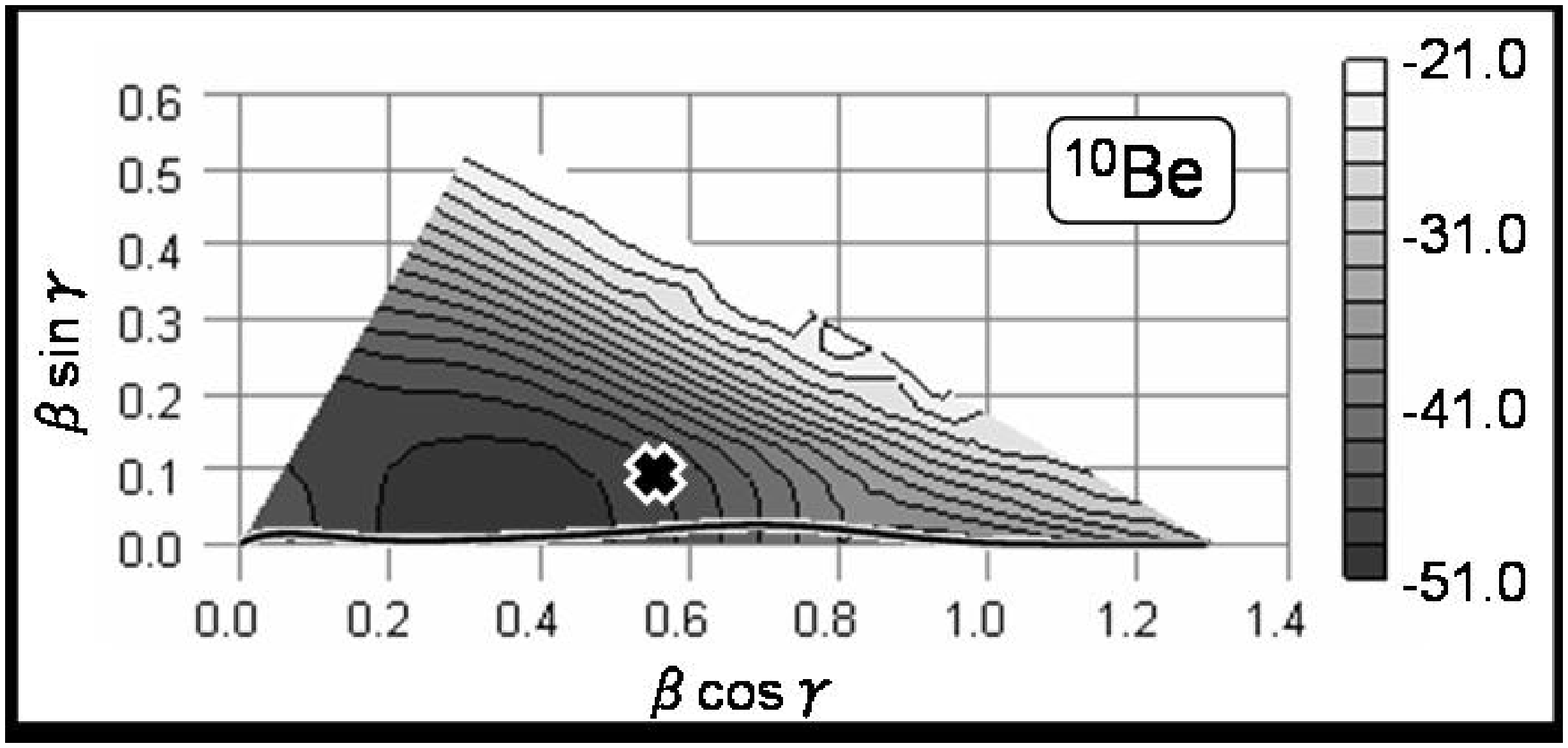}}
	\hfill
	\parbox{\halftext}{\includegraphics[width=6.6cm, bb=10 12 693 323, clip]{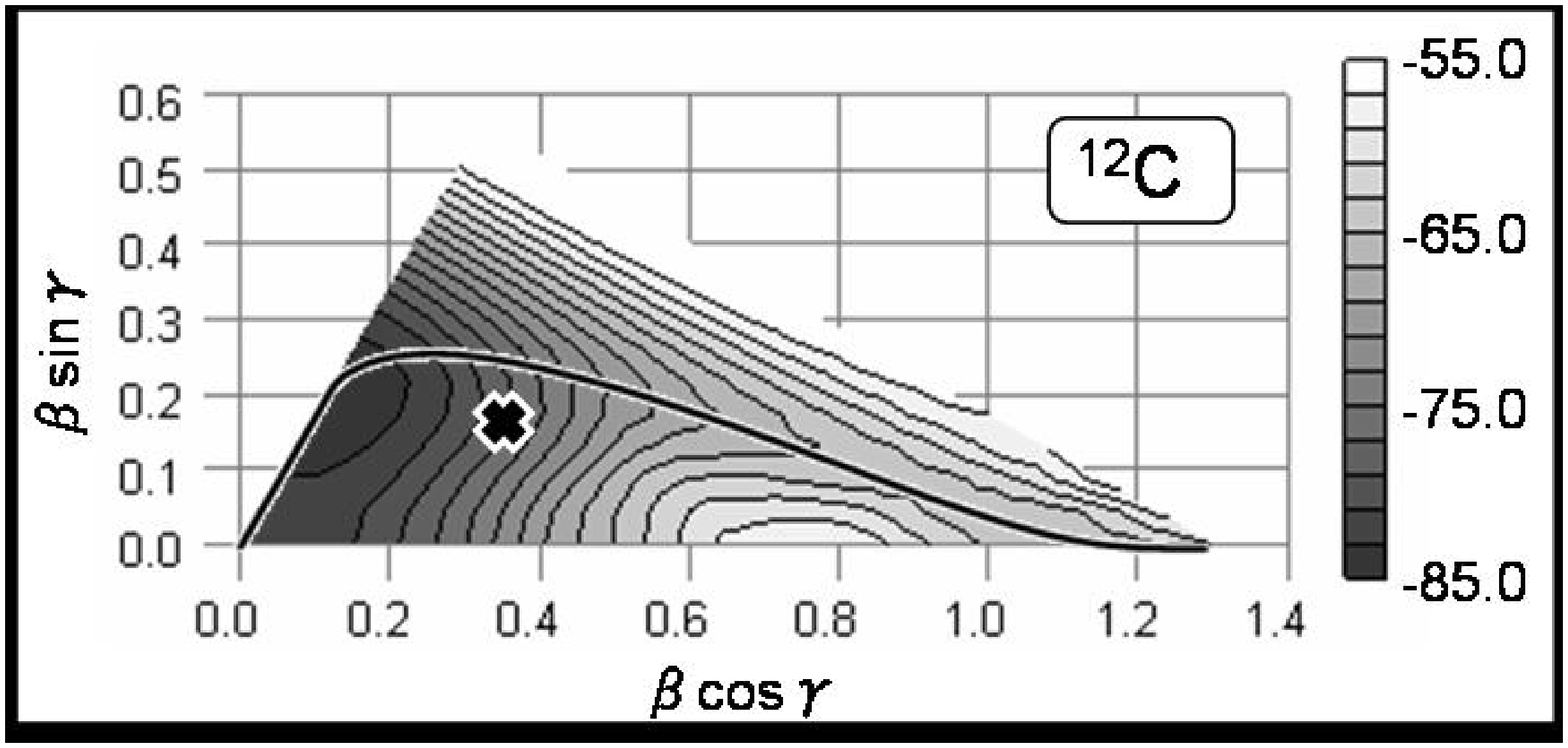}} \\
	\parbox{\halftext}{\includegraphics[width=6.6cm, bb=10 12 693 323, clip]{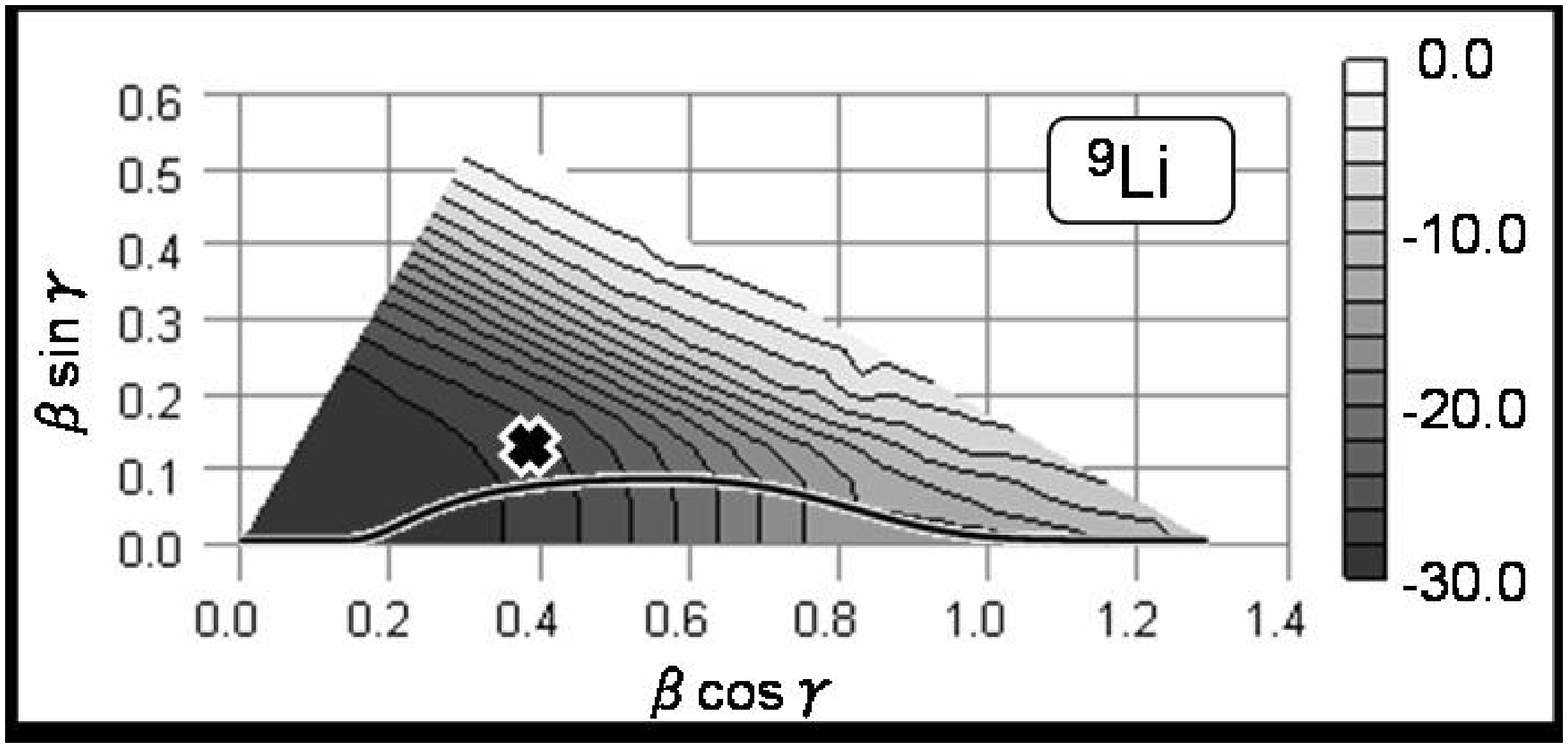}}
	\hfill
	\parbox{\halftext}{\includegraphics[width=6.6cm, bb=10 12 693 323, clip]{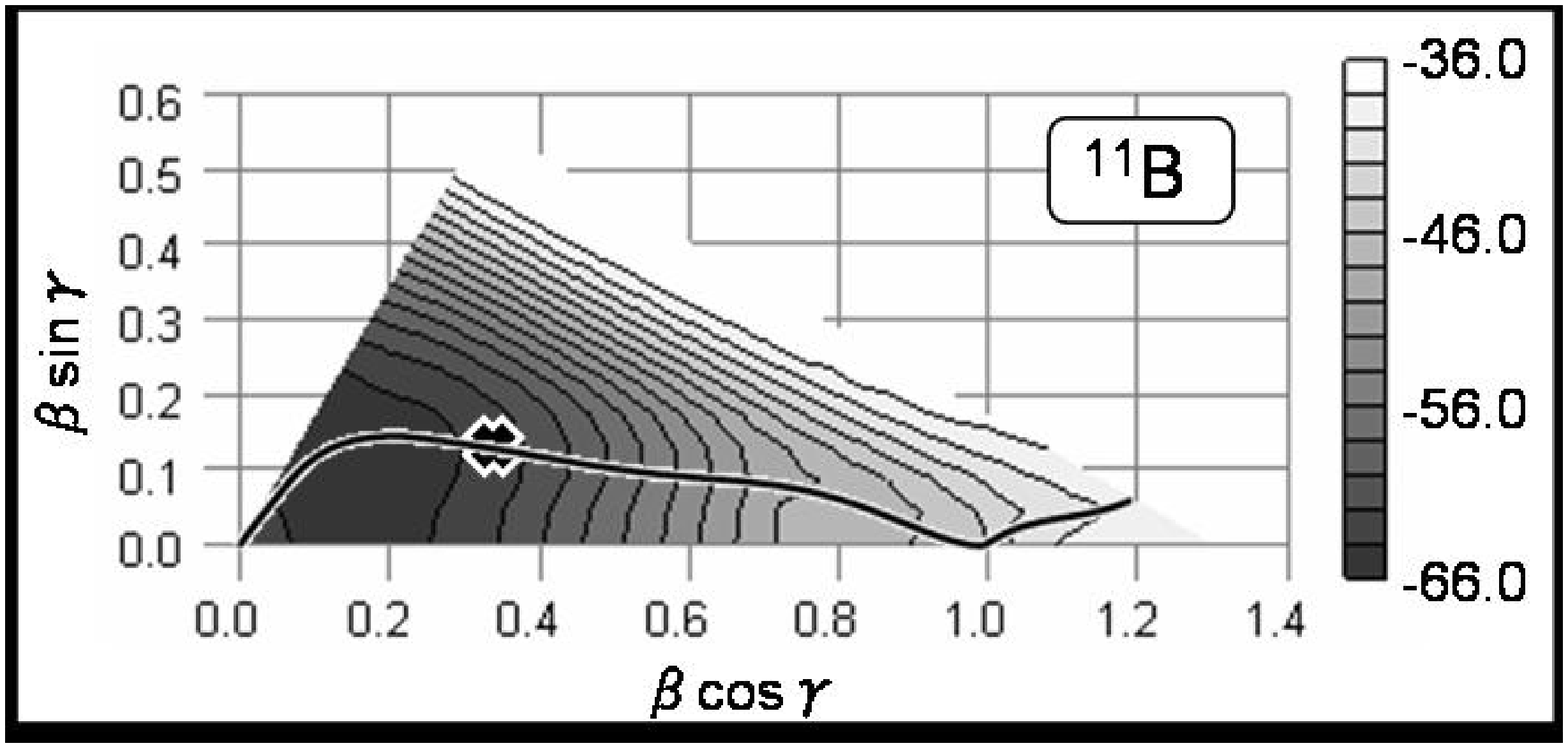}}
	\caption{$\beta$-$\gamma$ values obtained by the $\beta$ constraint AMD.
     		The solid lines show the $\beta$ constraint paths that indicate
            the optimum $\gamma$ of the energy minimum position for a given $\beta$ value.
            The energy surfaces before total-angular-momentum projection calculated by
            the $\beta$-$\gamma$ constraint AMD are also shown with the contour map.
			Cross points stand for the energy minimum after the total-angular-momentum projection
            obtained by the $\beta$-$\gamma$ constraint AMD.}
	\label{beta_const_line}
\end{figure}

\begin{figure}[tb]
	\centering
	\includegraphics[angle=-90,width=6cm]{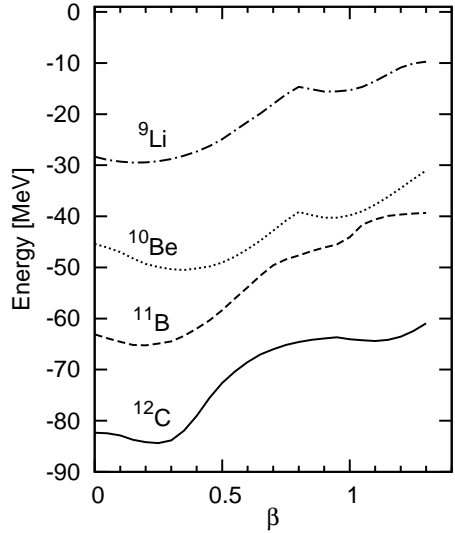}
	\caption{Energy curves calculated with $\beta$ constraint AMD.}
	\label{beta_constraint_energy_surface}
\end{figure}

Another method of the one-dimensional constraint is 
the $\beta$ constraint method without the axial symmetry assumption, which is  
usually applied to AMD calculations.
In this method, only the deformation parameter $\beta$ is constrained, and the triaxiality $\gamma$
is a free parameter determined in the energy variation.
Since $\gamma$ is automatically optimized for each $\beta$ constraint, it can take 
a nonzero value if a system favors a triaxial deformation.
This method has been found to be useful for
treating triaxial deformations of low-lying states.\cite{En'yo_AMD_03,Kimura_44Ti_06,Taniguchi_40Ca_07} \ 
Figure~\ref{beta_const_line} shows the paths of the $\beta$ constraint method 
on the $\beta$-$\gamma$ plane.
For reference, we also show the energy minimum point of the total-angular-momentum projected
energy surface with the cross points in the figure.
The energy curves along the paths obtained by the $\beta$ constraint method
are plotted as a function of $\beta$ in Fig.~\ref{beta_constraint_energy_surface}.
As seen in Fig.~\ref{beta_const_line}, the $\beta$ constraint path goes {\it by} the 
cross point for the energy minimum point of the total-angular-momentum projected energy surface
but does not necessarily pass {\it on} the minimum.
The difference in the minimum energy after the total-angular-momentum projection
between the $\beta$ constraint and $\beta$-$\gamma$ constraint is only 1 MeV
in cases of $^{10}$Be, $^{12}$C, and $^{9}$Li.
Therefore, we can say that the $\beta$ constraint method is reasonable, at least, 
in the description of the ground state properties.
However, as we explain below, the basis wave functions obtained by the 
$\beta$ constraint method are insufficient to describe some excited states of $^{10}$Be and
$^{12}$C, and the two-dimensional $\beta$-$\gamma$ constraint is found to be essential.

In the case of $^{10}$Be, the prolate deformations are obtained mostly by 
the $\beta$ constraint, as seen in Fig.~\ref{beta_const_line},
which shows that the $\beta$ constraint path goes along the $\gamma = 0^{\circ}$ axis.
In other words, the triaxially deformed structures in the finite $\gamma$ region cannot be
obtained within the $\beta$ constraint method.
On the other hand, the $2^{+}_{2}$ and $3^{+}_{1}$ states in the $K^{\pi}=2^{+}$ side band 
are obtained from the triaxially deformed basis wave functions.
That is, the $\beta$ constraint method is insufficient to reproduce
the $K^{\pi}=2^{+}$ side band in $^{10}$Be.

In the case of $^{12}$C, the two-dimensional $\beta$-$\gamma$ constraint method is
crucial for describing the $0^{+}_{2}$ state as follows.
We have shown the squared overlaps of the basis wave functions at each point $(\beta, \gamma)$ 
with the GCM wave functions for the $0^{+}$ states in Fig.~\ref{overlap_12C_0_+_2}.
The $0^{+}_{1}$ and $0^{+}_{3}$ states have large overlaps with the basis wave functions
in the $(\beta \cos \gamma, \beta \sin \gamma) \sim (0.35, 0.17)$ and 
$(\beta \cos \gamma, \beta \sin \gamma) = (1.13, 0.04)$ regions, respectively. 
The $\beta$ constraint line goes through these regions, and therefore,
the $0^{+}_{1}$ and $0^{+}_{3}$ states may be reproduced with the basis wave functions 
on the $\beta$ constraint line.
However, the $0^{+}_{2}$ state has a large overlap with the basis wave functions 
in the broad $\gamma$ area of the large $\beta$ region, which 
cannot be approached in the one-dimensional $\beta$ constraint method.
In this area, various configurations of the developed three $\alpha$ clusters appear
depending on $\beta$ and $\gamma$.
As explained earlier, the linear combination of many 3$\alpha$-cluster configurations is essential 
for describing the $0^{+}_{2}$ state.
These facts indicate that the $\beta$ constraint method is insufficient 
and the $\beta$-$\gamma$ constraint method is necessary to describe the $0^{+}_{2}$ state of $^{12}$C.

\section{Summary and outlook}\label{summary}

To describe various cluster and shell structures in light nuclei,
we proposed a new method of $\beta$-$\gamma$ constraint AMD.   
We applied this method to the $N=6$ isotones, $^{10}$Be, $^{12}$C, $^{9}$Li, and $^{11}$B.
In the energy variation in the $\beta$-$\gamma$ constraint AMD,
various structures appear depending on the two-dimensional constraint parameters, $\beta$ and $\gamma$.
In the small $\beta$ region, shell-model-like structures appear.
In particular, the neutron structure has the dominant $p_{3/2}$-shell closed
configurations with the spherical shape.
In the large $\beta$ region, cluster structures develop well.
As a function of the $\gamma$ parameter, various cluster configurations appear. 
That is, in the prolate region along the $\gamma = 0^{\circ}$ line,
two-body cluster structures or linear-chainlike structures 
develop well as the deformation parameter $\beta$ becomes large.
In the oblate region along the $\gamma = 60^{\circ}$ line, 
three-body cluster structures with equilateral-triangular or isosceles-triangular configurations appear.
In the triaxial region, various triangle configurations of the three-body clusters are obtained.

For $^{10}$Be and $^{12}$C, we superposed the basis AMD wave functions obtained 
by the $\beta$-$\gamma$ constraint method to calculate the energy spectra.
We compared the present results with the experimental data. The results reproduce well the experimental
spectra and are consistent with other theoretical studies. It was proved that 
the $\beta$-$\gamma$ constraint method is useful for describing the ground and excited states of these nuclei.
We also mentioned the advantages of the two-dimensional $\beta$-$\gamma$ constraint method over the 
one-dimensional constraint method.
It was confirmed that the degree of freedom of the triaxiality $\gamma$ in the two-dimensional $\beta$-$\gamma$ constraint
has considerable importance to describe such excited states
as the $K^{\pi}=2^{+}$ side band of $^{10}$Be and the $0^{+}_{2}$ state of $^{12}$C.

Moreover, in $^{9}$Li and $^{11}$B, cluster structures are expected to appear in the excited states 
from the similarity of the energy surface between $^{9}$Li and $^{10}$Be, and that between
$^{11}$B and $^{12}$C. Further investigations with the GCM calculations using basis wave functions on the $\beta$-$\gamma$ plane
are required to discuss the energy spectra of $^{9}$Li and $^{11}$B.

The present method proved to be very effective for describing various cluster and shell structures
in the ground and excited states of light nuclei.
We have applied the $\beta$-$\gamma$ constraint AMD to other nuclei such as $^{14}$C,
where shell-model-like features and cluster aspect are expected to coexist.
We will report the results in a future paper.

\section*{Acknowledgements}

The computational calculations of this work were performed  by supercomputers in YITP and KEK. 
This work was supported by the YIPQS program in YITP.
It was also supported by the Grant-in-Aid for the Global COE Program 
``The Next Generation of Physics, Spun from Universality and Emergence" 
from the Ministry of Education, Culture, Sports, Science and Technology (MEXT) of Japan,
and a Grant-in-Aid from JSPS.

\end{document}